\documentclass{LMCS}

\def\doi{9(2:14)2013}
\lmcsheading%
{\doi}
{1--34}
{}
{}
{Jul.~\phantom03, 2012}
{Jun.~28, 2013}
{}

\pdfoutput=1

\usepackage{enumerate}
\usepackage{hyperref}
\hypersetup{hidelinks=true}
\hypersetup{pdfdisplaydoctitle=true,pdfauthor=Samy Abbes,pdftitle=Markov two-components processes}

\usepackage[english]{babel}
\usepackage{bm}
\usepackage[all]{xy}

\theoremstyle{plain}
\newtheorem{propdef}[thm]{Proposition and definition}

\newcommand{\tcp}{two-components process}
\newcommand{\abtcp}{{\normalfont M2CP}}
\newcommand{\slgb}{\mbox{$\sigma$-alge}\-bra}

\newcommand{\tdefine}[1]{\textbf{#1}}
\newcommand{\FFF}{\mathfrak{F}}
\newcommand{\GGG}{\mathfrak{G}}
\newcommand{\RRR}{\mathbb{R}}
\newcommand{\seq}[2]{(#1_{#2})_{{#2}\geq0}}
\newcommand{\TT}{\mathcal{T}}
\newcommand{\II}{\mathcal{I}}
\newcommand{\EE}{\mathcal{E}}
\newcommand{\NN}{\mathbb{N}}
\newcommand{\VV}{\mathcal{V}}
\newcommand{\bpr}{\mathbf{P}}
\newcommand{\bprq}{\mathbf{Q}}
\newcommand{\pr}{\mathbb{P}}
\newcommand{\tq}{\;|\;}
\newcommand{\as}{\text{a.s.}}
\newcommand{\besp}{\mathbf{E}}
\newcommand{\un}[1]{\mathbf{1}_{\{#1\}}}

\renewcommand{\c}{\bm c}
\renewcommand{\d}{\bm d}
\newcommand{\betatilde}{\widetilde{\beta}}
\newcommand{\doublebullet}{\fbox{\makebox[.5em]{$\begin{array}{c} \bullet\\[-.7em]\bullet\end{array}$}}}
\newlength{\smalldim}
\begin{document}

\title{Markov Two-Components Processes}

\author[S.~Abbes]{Samy Abbes}	
\address{Universit\'e Paris Diderot/Laboratoire PPS\\B\^atiment Sophie
  Germain, avenue de France \phantom{75013} 75\,013 Paris, France
}%

\email{samy.abbes@univ-paris-diderot.fr}  



\ACMCCS{[{\bf Mathematics of computing}]: Probability and
  statistics---Stochastic processes---Markov processes; [{\bf Theory
      of computation}]: Models of computation---Concurrency---Parallel
  computing models}



\begin{abstract}
  \noindent
  We propose Markov two-components processes (M2CP) as a probabilistic
  model of asynchronous systems based on the trace semantics for
  concurrency. Considering an asynchronous system distributed over two
  sites, we introduce concepts and tools to manipulate random
  trajectories in an asynchronous framework: stopping times, an
  Asynchronous Strong Markov property, recurrent and transient states
  and irreducible components of asynchronous probabilistic
  processes. The asynchrony assumption implies that there is no global
  totally ordered clock ruling the system. Instead, time appears as
  \emph{partially ordered} and \emph{random}.

  We construct and characterize M2CP through a finite family of
  transition matrices. M2CP have a local independence property that
  guarantees that local components are independent in the
  probabilistic sense, conditionally to their synchronization
  constraints. A synchronization product of two Markov chains is
  introduced, as a natural example of M2CP.
\end{abstract}

\maketitle

\medskip

{\small\noindent\emph{Dedicated to the memory of Philippe Darondeau} (\oldstylenums{1948}--\oldstylenums{2013})}

\medskip

\section*{Introduction}
\label{sec:introduction}

\subsection*{General settings and requirements}
\label{sec:gener-sett-requ}

In this paper we introduce a probabilistic framework for a simple
asynchronous system distributed over two sites, based on the trace
semantics of concurrency. Consider a communicating system consisting
of two subsystems, called site $1$ and site~$2$, that need to
synchronize with one another from time to time, for example for
message exchange. Intended applications are, for instance, simple
client-server situations, device-device driver interactions,
communication bridge between two asynchronous networks.  The
synchronization is modeled, for each site, by the fact that the
concerned subsystem is entering some \emph{synchronizing state},
corresponding to a synchronization task---there shall be several
synchronization states corresponding to different tasks. It is natural
to consider that the synchronization states are shared: both
subsystems are supposed to enter together into a shared
synchronization state. Beside synchronization states, we assume that
each subsystem may evolve between other states that concern the local
activity of each subsystem, and seen as \emph{private states}.  Hence
we consider for each site $i=1,2$ some finite set of states~$S^i$,
with the intended feature that $Q=S^1\cap S^2$ is a nonempty set of
synchronization states.

Whenever the two subsystems enter some of their private states, the
corresponding events are said to be \emph{concurrent}. It is natural
to consider that the private activity of a given site should not
influence the private activity of the other site. This ought to be
reflected by some kind of statistical independence in the
probabilistic modeling.  Another feature that we are seeking is that
the \emph{local} time scales of private activities do not need to be
synchronous. Indeed, the local time scale of each subsystem might be
driven for instance by the input of a user, by the arrival of network
events, or by its internal chipset clock; therefore, it is realistic
\emph{not} to assume any correlation between local time scales, but
for synchronization. In particular, in a discrete time setting, the
synchronization instants counted on the two different local time
scales shall not need to be equal, making the two subsystems
\emph{asynchronous}.

\subsection*{Sequential probabilistic systems and concurrency}
\label{sec:limit-sequ-prob}

Classically, Markov chains in either discrete or continuous time are a
popular model for adding a probabilistic layer to describe the
evolution of a transition system.  Since the Markov chain formalism is
intrinsically sequential, its straightforward application to a
concurrent system brings the issue of translating a concurrent system
into a sequential one. A solution to this issue, found in the
Probabilistic Automata literature for instance
\cite{segala95,lynch03}, is the introduction of a non deterministic
scheduler in charge of deciding which subsystem is about to run at
each time instant. This defines a Markov Decision Process, a model
introduced earlier for control issues in~\cite{bellman57}.  Other ways
of composing probabilistic systems to form a Markov process, with or
without non determinism, are usually based on Milner's CCS
\cite{milner89} or Hoare's CSP~\cite{hoare85}, where the
synchronization policy for possibly synchronizing processes is either
to allow or to force synchronization. In \cite{argenio99} for
instance, where both synchronization methods \`a la CSS and \`a la CSP
are encoded in the model of bundle probabilistic transition systems,
renormalization occurs at each step to take into account the selected
synchronization paradigm.

\subsection*{Probabilistic trace semantics. Lattice of trajectories}
\label{sec:prob-trace-semant}

We introduce another way of randomizing our simple concurrent
system. We first accept as a basic fact that modeling the evolution of
a system as ordered paths of events jeopardizes the concurrency
feature of the model. Adopting instead the so-called trace semantics
for concurrency (or partial order semantics)~\cite{nie80,nie95},
lattices replace ordered paths to model trajectories. Unordered events
of a trajectory are then intrinsically concurrent. This raises a
question on the probabilistic side: which part of Markov chain theory
can we rebuild on this new basis?

The aim of this paper is to provide an answer to the question. Our
work is thus largely inspired by Markov chain theory; but we try to
adapt the theory to the partial order semantics of concurrency,
instead of directly turning a concurrent system into a Markov chain
(or a variant of it) as in~\cite{lynch03,argenio99}.

Let us be precise about what we mean in this paper by a partial order
semantics for concurrency, referring to the two sets of local states
$S^1$ and~$S^2$ with synchronization constraint $Q=S^1\cap S^2$. We
will then explain how probability concepts apply in this setting.

If two sequences of states in $S^1\cup S^2$ only differ by the
interleaving order of private states of different sites, such as
$a\cdot e$ and $e\cdot a$ with $a\in S^1\setminus Q$ and $e\in
S^2\setminus Q$, the trace semantics suggests to simply identify them:
$a\cdot e\equiv e\cdot a$.  Propagating this identification to
sequences of events of arbitrary length, we obtain an equivalence
relation on sequences. Sequences that cannot be permuted are those of
the form $x\cdot y$ with $x,y\in S^1$ or $x,y\in S^2$, which include
those of the form $x\cdot \c$ with $\c\in Q$ and any $x\in S^1\cup
S^2$. We adopt a simple representation for equivalence classes of
sequences by mapping each equivalence class to a \emph{pair} of
sequences, where each coordinate is reserved for a given site;
synchronization states appear in both coordinates. Hence the
equivalence class of $a\cdot e\equiv e\cdot a$ is mapped to $(a,e)$,
the equivalence class of $a\cdot e\cdot\c\equiv e\cdot a\cdot \c$ is
mapped to $(a\cdot\c,e\cdot \c)$. We define thus a \emph{trajectory}
as a pair $(s^1,s^2)$, where $s^i$ is a sequence of elements
in~$S^i$\,, and such that the sequences of synchronization states
extracted from $s^1$ and from~$s^2$, and taken in their order of
appearance, shall be \emph{equal}.

An \emph{infinite trajectory} is defined as a trajectory
$\omega=(\omega^1,\omega^2)$ where both sequences $\omega^1$ and
$\omega^2$ are infinite. So for example, if $S^1=\{a, b, \bm c, \bm
d\}$ and $S^2=\{\bm c,\bm d,e,f\}$, and thus \mbox{$Q=\{\bm c,\bm
  d\}$}, an infinite trajectory could be $\omega=(\omega^1,\omega^2)$
with $\omega^1$ and $\omega^2$ starting as follows:
\mbox{$\omega^1=a\cdot \bm c\cdot b\cdot a\cdot b\cdot b\cdot \bm
  d\cdot(\cdots)$} and \mbox{$\omega^2=e\cdot f\cdot \bm c\cdot f\cdot
  \bm d\cdot(\cdots)$}. The common extracted sequence of
synchronization states starts in this example with $\bm c\cdot\bm d$.
Note the important feature that each local trajectory $\omega^i$ is
permitted to have a free evolution between synchronizations:
synchronizations occur at instants $2$ and $7$ for~$\omega^1$, while
they occur at instants $3$ and $5$ for~$\omega^2$; here, the instants
of synchronization are relative to the \emph{local} time scales. The
set $\Omega$ of infinite trajectories is the natural sample space to
put a probability measure on.

\begin{figure}
\centerline{ \xymatrix{&(a,\epsilon)\ar[dr]\\
(\epsilon,\epsilon)\ar[ur]\ar[dr]&&(a,e)\ar[r]&(a,e\cdot
f)\ar[r]&(a\cdot\c,e\cdot f\cdot\c)
\\
&(\epsilon,e)\ar[ur]\ar[r]&(\epsilon,e\cdot f)\ar[ur]
}}
  \caption{\textsl{Lattice of subtrajectories of
      $v=(a\cdot\c,e\cdot f\cdot\c)$.}}
  \label{fig:qsdqsjpa}
\end{figure}
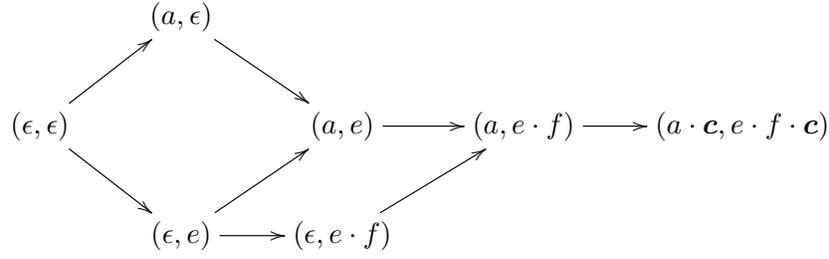

There is a natural notion of \emph{subtrajectory}: in the previous
example, $v=(a\cdot\c,e\cdot f\cdot\c)$ is a finite subtrajectory of
$\omega=(\omega^1,\omega^2)$. ``Being a subtrajectory'' defines a
binary relation that equips subtrajectories of a given trajectory with
a \emph{lattice} structure. For instance, and denoting by $\epsilon$
the empty word, the subtrajectories of $v$ are: $(\epsilon,\epsilon)$,
$(a,\epsilon)$, $(\epsilon,e)$, $(\epsilon,e\cdot f)$, $(a,e)$,
$(a,e\cdot f)$ and $(a\cdot\c,e\cdot f\cdot\c)$. Their lattice is
depicted in Figure~\ref{fig:qsdqsjpa}. Observe that, for a given
trajectory, its subtrajectories are naturally identified with
two-components ``time instants''. In case of~$v$, these time instants
are $(0,0)$, $(1,0)$, $(0,1)$, $(0,2)$, $(1,1)$, $(1,2)$ and~$(2,3)$,
and they form a sublattice of the lattice $\NN\times \NN$. However,
even if one considers an infinite trajectory~$\omega$, the associated
lattice of two-components time instants is only a \emph{sub}lattice of
$\NN\times\NN$ in general. For instance, if $\zeta$ is any infinite
trajectory that has $v$ as subtrajectory, then $(2,2)$ is a time
instant that does not correspond to any subtrajectory of~$\zeta$,
because of the synchronization on state~$\c$.

Obviously, considering another trajectory $\omega'$ would lead to
another lattice of subtrajectories, not necessarily isomorphic to the
one associated with~$\omega$. We sum up the previous observations by
saying that time is \emph{partially ordered} on the one hand, since
time instants form a lattice and not a total order, and \emph{random}
on the other hand, since the lattice structure depends on the
trajectory considered, that is, on the execution of the system.

\bigskip\bigskip

\subsection*{Defining M2CP: absence of transition matrix}
\label{sec:absence-trans-matr}

This has consequences for the way one may construct a probability
measure on the space $\Omega$ of infinite trajectories.  Consider
again the finite trajectory encountered above, $v=(a\cdot\c,e\cdot
f\cdot\c)$. The occurrences of $a$ on site~$1$, and of $e$ on
site~$2$, are \emph{concurrent}. Trying to determine the precise
interleaving of $a$ and $e$ is irrelevant for us. This desired feature
prevents us from applying the standard recursive construction to
assign a probability to trajectory~$v$ (that is: the probability that
$v$ occurs as a subtrajectory of a sample infinite
trajectory~$\omega$): starting from the initial state, there is no
obvious choice between $a$ and~$e$; which one should be first plugged
in the probability computation?

Therefore the lattice structure of trajectories implies to give up, at
least temporarily, the familiar inductive computation of probabilities
based on transition matrices. Nevertheless, two important notions can
be defined in the asynchronous framework by analogy with Markov chain
theory: first, the notion of state reached by (``after'') a finite
trajectory (\S~\ref{sec:general-framework-1}); second, the
probabilistic evolution of the system ``after'' execution of a finite
trajectory (Definition~\ref{def:2} in \S~\ref{sec:prob-two-comp}). We define
a \emph{Markov two-components process} (\abtcp) as a random system
where the probabilistic future after execution of a finite trajectory
$v$ only depends on the state reached by~$v$.

\bigskip\bigskip

\subsection*{Stopping times for M2CP}
\label{sec:stopping-times-m2cp}

Recall that a stopping time in Markov chain theory identifies with a
random halting procedure that does not need anticipation: an observer
can decide whether the stopping time has been reached based on the
only knowledge of the process history at each step. Stopping times are
a basic tool in Markov chain theory. Important notions such as the
first return time to a state, recurrent and transient states are
defined by means of stopping times.  Stopping times are manipulated
with the help of the Strong Markov property, a central result in
Markov chain theory. We show that several aspects of the Markovian
language carry over to the asynchronous framework. Once an adequate
notion of stopping time for asynchronous probabilistic processes has
been defined (Definition~\ref{def:6} in~\S~\ref{sec:stopping-times}),
derived notions such as the first hitting time to a state, and the
notions of recurrent and transient states follow by almost literally
translating the original ones into the asynchronous language. We show
that the Strong Markov property also has an equivalent, called the
Asynchronous Strong Markov property, which serves as a basic tool for
probabilistic reasoning. Some other notions translate in a more subtle
way: the first reaching time of a \emph{set} of states needs some
additional care, since the lattice structure of trajectories prevents
a straightforward generalization of the analogous notion from Markov
chain theory, providing an interesting difference with Markov chain
theory. Irreducible processes have an equivalent counterpart in the
asynchronous framework, and we detail the decomposition of a \abtcp\
into irreducible components.

\subsection*{The Local Independence Property}
\label{sec:local-indep-prop}

Therefore, we have on the one hand these notions obtained as a
generalization of analogous notions from Markov chain theory to the
asynchronous framework.  But on the other hand, we also have other
notions specific to the asynchronous framework, and that would not
make sense for Markov chains. In particular,
%
%
the way the two local
components behave with respect to one another is a question specific
to the asynchronous framework. Since the two local components
synchronize with one another, they cannot be fully independent in the
probabilistic sense. There is however a weaker notion of independence
in probability theory, adapted to our purpose, which is
\emph{conditional} independence. We call Local Independence Property
(LIP) the property that the two components are independent
conditionally to their synchronization constraint. Informally, the LIP
says that the local components have the maximal independence that they
can have, up to their synchronization constraint. We characterize
\abtcp\ with the LIP by a finite family of transition matrices; and we
show how to construct a \abtcp\ from an adapted family of such
transition matrices.  The finite collection of numbers this family of
matrices defines is an equivalent, in the asynchronous framework, of
the transition matrix for a Markov chain.


\subsection*{Synchronization of systems}
\label{sec:synchr-syst}

The composition of probabilistic systems has always been a challenge,
with multiple applications in the theory of network analysis
\cite{argenio99,lynch03,baccelli92:_synch_linear}. The main limitation
of the theory of probabilistic event structures as it has been
developed so far by the author together with A.~Benveniste
in~\cite{abbes08,abbes06a} (another probabilistic model with trace
semantics targeting applications to probabilistic $1$-safe Petri
nets), and by other authors in \cite{varacca04:_probab} is the non
ability to define a suitable synchronization product. This very
limitation has motivated the development of the present framework, by
starting with the definition of a synchronization product for two
Markov chains. By recursively ``forcing'' their synchronization, it is
shown in this paper how the synchronization of two Markov chains on
shared common states naturally leads to a \abtcp. Even if one was
interested in this construction only (the author is aware of current
work on this kind of \textit{a priori} model, simply because it was
the only one people could think of), including it inside a more
general picture as it is done in this paper is useful to better
understand its properties.

\subsection*{Organization of the paper}
\label{sec:organization-paper}

We describe the model in~\S~\ref{sec:prob-proc-mark}, defining a
general notion of probabilistic \tcp, and then specializing to
\emph{Markov} \tcp es. In~\S~\ref{sec:synchr-two-mark} we introduce
the synchronization product of two Markov chains. This construction
provides an example of \abtcp, intended to support the intuition for
\abtcp\ in general. Next section, \S~\ref{sec:stopp-times-strong}, is
devoted to Markovian concepts in the asynchronous framework, centered
around the Asynchronous Strong Markov property. We introduce
recurrence and transience of states and the decomposition of \abtcp\
into irreducible components. The new notions of closed and open
processes are studied in this section, as well as the definition of
stopping times for asynchronous processes. The Local Independence
Property (LIP) is the topic of~\S~\ref{sec:prop-mark-two}, and it is
shown that the synchronization of Markov chains introduced in
\S~\ref{sec:synchr-two-mark} satisfies the LIP. Finally,
\S~\ref{sec:gener-constr-mark}~is devoted to the construction and
characterization of general \abtcp\ with the LIP. A concluding section
introduces directions for future work. It discusses limitations
imposed by the two-components hypothesis, and possible ways to remove
this limitating hypothesis.

\section{Probabilistic Processes and 
Markov Processes on Two Sites}
\label{sec:prob-proc-mark}

\subsection{General Framework}
\label{sec:general-framework-1}

A \tdefine{distributed system} is given by a pair $(S^1,S^2)$, where
$S^i$ for $i=1,2$ is a finite set, called the set of \tdefine{local
  states} of site~$i$.  A \tdefine{local trajectory} attached to site
$i$ is a sequence of local states of this site. For $i=1,2$, we denote
by $\Omega^i$ the set of infinite local trajectories attached to
site~$i$.

The two local state sets $S^1$ and $S^2$ are intended to have a non
empty intersection, otherwise the theory has little interest. We put
$Q=S^1\cap S^2$. Elements of $Q$ are called \tdefine{common states} or
\tdefine{shared states}. In contrast, states in $S^i\setminus Q$
are said to be \tdefine{private to site~$\bm i$}, for $i=1,2$. From
now on, \textbf{we will always assume that $\bm S^{\bm
    i}\bm\setminus\bm Q\bm\neq\bm\emptyset$} for $i=1,2$: each site
has at least one private state. This is a convenient technical
assumption; removing it would not harm if needed.

Given a sequence $(x_j)_j$ of elements in a set~$S$, either finite or
infinite, and given a subset $A\subseteq S$, the \tdefine{$\bm
  A$-sequence induced by $\bm(\bm x_{\bm j}\bm)_{\bm j}$} is defined
as the sequence of elements of $A$ encountered by the
sequence~$(x_j)_j$\,, in their order of appearance. Given two local
trajectories $\seq {x^1}n$ and $\seq{x^2}n$ on sites $1$ and $2$
respectively, we will say that they \tdefine{synchronize} if the two
$Q$-sequences they induce are \emph{equal}. A pair of two
synchronizing local trajectories will be called \tdefine{a global
  trajectory}, or simply a \tdefine{trajectory} for brevity. Among
them, \tdefine{finite trajectories} are those whose components are
both finite sequences of states.

Trajectories are ordered component by component: if $s=(s^1,s^2)$ and
$t=(t^1,t^2)$ are two trajectories, we define $s\leq t$ if $s^1\leq
t^1$ and $s^2\leq t^2$, where the order on sequences is the usual
prefix order. The resulting binary relation on trajectories is a
partial order, the maximal elements of which are exactly those whose
components are both infinite: this relies on the fact that
$S^i\setminus Q\neq\emptyset$ for $i=1,2$ (for instance, if
$S^1=\{a,\bm b\}$ and $S^2=\{\bm b\}$ so that $Q=\{\bm b\}$ and
$S^2\setminus Q=\emptyset$, then $(\bm baaa\cdots,\bm b)$ is maximal,
but the second component is finite).  The set of maximal trajectories
is denoted by~$\Omega$, and we have that
$\Omega\subseteq\Omega^1\times\Omega^2$. For $s$ a \emph{finite}
trajectory, the subset of $\Omega$ defined by
\begin{equation}
\label{eq:11}
\uparrow s=\{\omega\in\Omega\tq s\leq\omega\}
\end{equation}
is called the \tdefine{elementary cylinder} of base~$s$---adapting a
standard notion from Measure theory to our framework. 

Given any trajectory $s=(s^1,s^2)$, the \tdefine{subtrajectories
  of~$\bm s$} are those trajectories $t$ such that $t\leq s$. Observe
that not any prefix $t$ of $s$ is a subtrajectory; since $t$ could
very well not be a trajectory itself.

Given a trajectory $(s^1,s^2)$, we denote by $(y_j)_j$ the
$Q$-sequence induced by both sequences $s^1$ and~$s^2$\,. It can be
finite or infinite, even empty. We refer to $(y_j)_j$ as to the
\tdefine{$\bm Q$-sequence induced} by $ (s^1,s^2)$.

A \tdefine{global state} is any pair $\alpha=(x^1,x^2)\in S^1\times
S^2$. We reserve the letters $\alpha$ and $\beta$ to denote global
states. Observe that trajectories are \emph{not} defined as sequences
of global states; since the length of the two components may very well
differ. Let $\alpha=(x,y)$ be some fixed global state, thought of as
the \emph{initial} state of the system.  If $s=(s^1,s^2)$ is a finite
trajectory, we define
\begin{equation}
\label{eq:2}
\gamma_\alpha(s)=(x^1,x^2)\in S^1\times S^2
\end{equation}
as the pair of last states of the two sequences $x\cdot s^1$
and~$y\cdot s^2$\,. We understand $\gamma_\alpha(s)$ as the current
global state after the execution of finite trajectory~$s$,
starting from~$\alpha$. Note that, with this definition,
$\gamma_\alpha$~is well defined on the empty sequence and
$\gamma_\alpha(\emptyset)=\alpha$. By an abuse of notation, we will
omit $\alpha$ and write $\gamma$ instead of~$\gamma_\alpha$\,, the
context making clear which initial state $\alpha$ we refer to.

We introduce a notion of length for trajectories. We denote by
$\TT$ the set
\begin{equation*}
  \TT=\bigl(\NN\times\NN)\cup\{\infty\}\,.
\end{equation*}
The set $\TT$ is partially ordered component by component, with the
natural order on each component, and $(m,n)\leq\infty$ for all
$(m,n)\in\NN\times\NN$.  If $s=(s^1,s^2)$ is any trajectory, the
\textbf{length of~$\bm s$} is defined by
\begin{equation*}
  |s|=
  \begin{cases}
(|s^1|,|s^2|)\in\TT,&\text{if $s$ is finite,}\\
\infty,&\text{otherwise,}    
  \end{cases}
\end{equation*}
where $|s^1|$ and $|s^2|$ denote the length of sequences. Roughly
speaking, lengths can be thought of as time instants; it becomes then
clear that time is only partially ordered, and not totally
ordered---see random times in \S~\ref{sec:stopping-times}
for a finer notion.

There is a \tdefine{concatenation} operation partially defined on
trajectories. If $s=(s^1,s^2)$ is a \emph{finite trajectory}, and
$t=(t^1,t^2)$ is any trajectory, then the concatenation denoted by
$s\cdot t$ and defined by $s\cdot t=(s^1\cdot t^1,s^2\cdot t^2)$ is
obviously a trajectory. If $t\in\Omega$, then $s\cdot t\in\Omega$ as
well. There is an obvious \tdefine{addition} on lengths, compatible
with concatenation of finite trajectories, in the sense that $|s\cdot
t|=|s|+|t|$.  If we fix~$s$, the concatenation
defines a bijection onto the cylinder of base~$s$:
\begin{equation}
\label{eq:1}
\Phi_s:
\begin{cases}
  \Omega\to\uparrow s\\
\omega\mapsto\Phi_s(\omega)= s\cdot \omega\,.
\end{cases}
\end{equation}

\subsection{Trajectory  Structure}
\label{sec:struct-traj}

The fact that we consider only two sites allows to precisely describe
the structure of trajectories. 

\begin{defi}
  \label{def:5}
 \
 \begin{enumerate}[(1)]
  \item An \tdefine{elementary trajectory} is a finite trajectory with
    a unique synchronization, that occurs at its end. Equivalently, a
    finite trajectory $s$ is elementary if $\gamma(s)=(x,x)$
    for some $x\in Q$, and $(x)$ is the $Q$-sequence induced by~$s$.
  \item We say that a trajectory is \tdefine{synchronization free} if
    its associated $Q$-sequence is empty.
  \end{enumerate}
\end{defi}

\noindent We omit the proof of the following proposition, which is elementary,
but fundamental for some constructions introduced later
in~\S~\ref{sec:synchr-two-mark} and
in~\S~\ref{sec:gener-constr-mark}.

\begin{prop}
  \label{prop:1}
\
  \begin{enumerate}[\em(1)]
  \item\label{item:10} Any finite trajectory has a unique
    decomposition as a concatenation of elementary trajectories,
    followed by a synchronization free trajectory.
  \item\label{item:11} Any maximal trajectory is either, according to
    its $Q$-sequence being infinite or finite:
    \begin{enumerate}[\em(a)]
    \item\label{item:1} A countable infinite concatenation of elementary
      trajectories, and the decomposition as such a concatenation is
      unique; or
    \item\label{item:2} A finite concatenation of elementary
      trajectories, followed by a synchronization free trajectory,
      infinite on both sides. This decomposition
      is unique.
    \end{enumerate}
  \end{enumerate}
\end{prop}

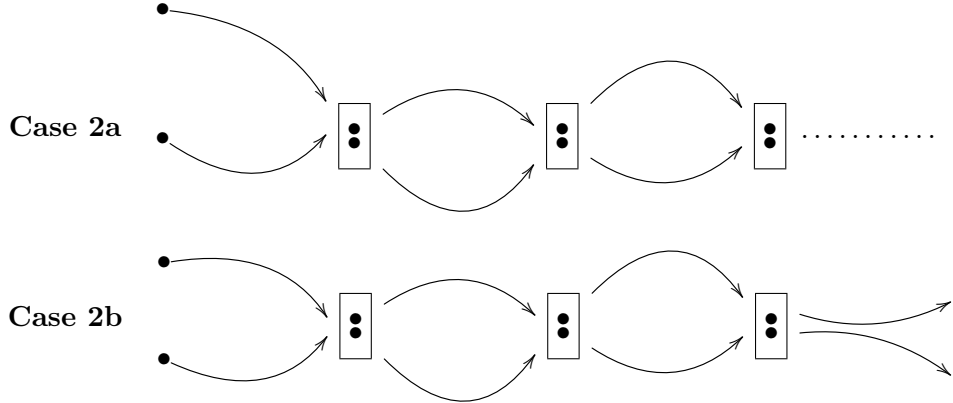
\begin{figure}
  \centering
\centering
\mbox{\textbf{Case~2a}\quad\xymatrix@1@C=2cm@M=1ex@R=3em{
{\rule{0em}{2em}}\POS!U(2)\drop{\bullet}\ar@/^2em/[r]!D(1.5)
\POS!D(20)\drop{\bullet}
\ar@/_2em/[r]!U(1)
&
\doublebullet
\ar@/^2em/[r]!D(.5)
\ar@/_3em/[r]!U(.5)
&
\doublebullet
\ar@/^3em/[r]!D(.5)
\ar@/_2em/[r]!U(.5)
&
\doublebullet
&\rule{1em}{0em}\POS!L(4)\drop{\makebox[2cm]{\dotfill}}
}}\par\bigskip\bigskip\bigskip
\mbox{\textbf{Case~2b}\quad\xymatrix@1@C=2cm@M=1ex@R=3em{
{\rule{0em}{2em}}\POS!U(1)\drop{\bullet}\ar@/^2em/[r]!D(1.5)
\POS!D(15)\drop{\bullet}
\ar@/_2em/[r]!U(1)
&
\doublebullet
\ar@/^2em/[r]!D(.5)
\ar@/_3em/[r]!U(.5)
&
\doublebullet
\ar@/^3em/[r]!D(.5)
\ar@/_2em/[r]!U(.5)
&
\doublebullet\POS!U(.5)\ar@/_1em/[r]!U(4.5)
\POS!D(.5)\ar@/^1em/[r]!D(4.5)
&\rule{1em}{0em}
}}%
\par\vspace{2em}
\caption{\textsl{Illustration of the decomposition of a maximal
    trajectory, according to Proposition~{\normalfont\ref{prop:1}},
    Cases~{\normalfont\ref{item:1}} and~{\normalfont\ref{item:2}}. The
    framed boxes represent the synchronizations, the arrows represent
    the private paths. In Case~{\normalfont\ref{item:1}}, the synchronization
    pattern keeps repeating on the right.}}
\label{fig:gt}
\end{figure}

\noindent Figure~\ref{fig:gt} depicts the decomposition of global trajectories
in cases~\ref{item:1} and~\ref{item:2}.  Finally, the following lemma
will be useful.

\begin{lem}
  \label{lem:5}
  For any trajectory $v$, the set of subtrajectories of $v$ is a well
  founded and complete lattice. Lower and upper bounds are taken
  component by component.
\end{lem}

\proof
  Let $v=(s^1,s^2)$, and let $\II(s^i)$ denote, for $i=1,2$, the set
  of initial subsequences of~$s^i$\,.  It is well known that
  $\II(s^i)$ is a total and well-founded order with arbitrary
  \emph{lub}s (least upper bounds). Therefore the component-wise order
  on $\II(s^1)\times\II(s^2)$ is a complete lattice, with lower and
  upper bounds taken component by component.

  To prove the lemma, it suffices thus to check that the
  component-wise upper and lower bounds of subtrajectories of $v$
  yield again subtrajectories of~$v$, and this is obvious, hence we
  are done.\qed

\subsection{Probabilistic Two-Components Processes}
\label{sec:prob-two-comp}

Although time has been abstracted from the framework, the notion of
trajectory is still present; this is all we need to introduce a
probabilistic layer. We consider the \slgb\ $\FFF$ on $\Omega$
generated by the countable family of elementary cylinders, defined
above in Eq.~(\ref{eq:11}). The \slgb\ $\FFF$ coincides with the trace
on $\Omega$ of the product \slgb\ on the infinite product
$\Omega^1\times\Omega^2=(S^1\times S^2)^\NN$\,, where of course~$S^i$,
as a finite set for $i=1,2$, is equipped with the discrete \slgb.

Unless stated otherwise, the set $\Omega$ will be equipped with the
\slgb~$\FFF$. Assume thus that $\bpr$ is a probability defined
on~$\Omega$. By an abuse of notation, if $s$ is a finite trajectory we
simply denote by $\bpr(s)$ the probability of the elementary cylinder
of base~$s$, so that $\bpr(s)=\bpr(\uparrow s)$. We say that a global
state $\alpha$ is \tdefine{reachable} w.r.t.~$\bpr$ if there exists a
finite trajectory $s$ such that $\bpr(s)>0$ and $\alpha=\gamma(s)$. A
probabilistic \tcp\ on a distributed system is defined as follows.

\begin{defi}
  \label{def:1}\
  \begin{enumerate}[(1)]
  \item A \tdefine{probabilistic two-components process}, or
    \tdefine{probabilistic process} for brevity, is a family
    $\pr=(\bpr_\alpha)_{\alpha\in X_0}$ of probability measures on
    $\Omega$ indexed by a set $X_0$ of global states, and satisfying
    the following property: for all \mbox{$\alpha\in X_0$}, if $\beta$
    is reachable with respect to~$\bpr_\alpha$\,, then \mbox{$\beta\in
      X_0$}\,.
  \item If $\beta$ is reachable w.r.t.~$\bpr_\alpha$\,, we say that
    $\beta$ is \tdefine{reachable from~$\bm \alpha$}.
  \item A \tdefine{subprocess} of a probabilistic process $\pr$ is a
    subfamily $(\bpr_\alpha)_{\alpha\in X_1}$\,, with $X_1\subset
    X_0$, that forms a probabilistic process.
  \end{enumerate}
\end{defi}

\noindent The probability $\bpr_\alpha$ is intended to describe the
probabilistic behavior of the system starting from~$\alpha$. However,
for technical reasons that will appear later, we consider the
evolution of the system \emph{after}~$\alpha$. In other words, we
assume that $\alpha$ has already been reached, and we put ourselves
just after it. In particular, we do \emph{not} assume that
$\bpr_\alpha(\uparrow\alpha)=1$, contrary to the usual convention
adopted in Markov chain theory.

\begin{defi}
  \label{def:17}
Let\/ $\pr=(\bpr_\alpha)_{\alpha\in X_0}$ be a probabilistic \tcp. Let
also $*$ be an arbitrary specified value not in $S^1\cup S^2$. 
\begin{enumerate}[(1)]
\item For $\omega\in\Omega$, we denote by
  $Y(\omega)=\bigl(Y_n(\omega)\bigr)_{n\geq0}$ the $Q$-sequence
  induced by~$\omega$, followed by the constant value $*$ if the
  $Q$-sequence is finite. In all cases, we also put $Y_{-1}=*$.  We
  refer to $Y$ as to the \tdefine{(random) synchronization sequence.}
\item We say that $\omega\in\Omega$ \tdefine{synchronizes infinitely often} if
  $Y_n(\omega)\neq *$ for all $n\geq0$.
\item We say that $\pr$ is \tdefine{closed} if for all $\alpha\in
  X_0$\,,  $Y_n\neq*$ for all
  $n\geq0$ and\/ $\bpr_\alpha$-\as
\item We say that\/ $\pr$ is \tdefine{open} if for all $\alpha\in
  X_0$\,,  $Y_n=*$ for all $n\geq0$ and\/
  $\bpr_\alpha$-\as
\end{enumerate}
\end{defi}

\medskip\noindent Consider any probability measure $\bpr$ on~$\Omega$, and let
$s$ be a finite trajectory. Observe that $\Phi_s:\Omega\to\uparrow s$
is not only a bijection, it is also bi-measurable. Considering the
action of $\Phi_s^{-1}$ on measures is thus meaningful. In particular,
if $\bpr(s)>0$, we define the probability $\bpr_s$ on $\Omega$ as the
image of the conditional probability $\bpr(\,\cdot\,|\uparrow s)$. It
satisfies, and is characterized by the relations
$\bpr_s(t)=\frac1{\bpr(s)}\bpr(s\cdot t)$, for $t$ ranging over the
set of finite trajectories.

\begin{defi}
  \label{def:2}
  If\/ $\bpr$ is a probability measure on\/~$\Omega$, and if $s$ is
  a finite trajectory such that\/ $\bpr(s)>0$, we define the probability
  measure $\bpr_s$ on\/ $\Omega$ characterized by:
\begin{equation}
  \label{eq:3}
\bpr_s(t)=\frac1{\bpr(s)}\bpr(s\cdot t),
\end{equation}
for $t$ ranging over the set of finite trajectories, as the
\tdefine{probabilistic future} of $s$ w.r.t. probability\/~$\bpr$.
\end{defi}

Markov \tcp es can now be defined as follows, without reference to any
explicit notion of time.

\begin{defi}
  \label{def:3}
  Given a distributed system, a \tdefine{Markov two-components
    process}, abbreviated\/ \tdefine{\abtcp}, is defined as a
  probabilistic process\/ $\pr=(\bpr_\alpha)_{\alpha\in X_0}$ over
  this system, satisfying the following property: for $\alpha$ ranging
  over $X_0$ and $s$ ranging over the set of finite trajectories such
  that\/ $\bpr_\alpha(s)>0$, the probabilistic future of trajectory
  $s$ w.r.t.~$\bpr_\alpha$ only depends on~$\gamma(s)$. This is
  equivalent to saying:
\begin{equation}
\label{eq:4}
\forall\alpha\in X_0\quad \forall s\quad \bpr_\alpha(s)>0\Rightarrow
\bigl(\bpr_\alpha\bigr)_s=\bpr_{\gamma(s)}\,.
\end{equation}
\end{defi}

Equation~\eqref{eq:4} formalizes the intuition that ``the
probabilistic future only depends on the present state''; we shall
refer to it as to the \emph{Markov property}. Some additional comments
about Definition~\ref{def:3}:
\begin{enumerate}[1.]
\item Markov chains are usually defined by their transition matrix,
  from which a probability measure on the space of trajectories is
  derived. Here, on the contrary, the lack of a totally ordered time
  index leads us to first consider a measure on the space of
  trajectories with the Markov property already encoded in it. It will
  be our task to find an equivalent for the transition matrix,
  that would characterize the probability measure through a finite
  number of real parameters with adequate normalization
  conditions. This is the topic of~\S~\ref{sec:gener-constr-mark}.
\item Considering the same definition for a probability measure on a
  space of trajectories with only one component---for instance, taking
  $S^2=\{*\}$ a singleton disjoint from~$S^1$---, would
  exactly bring us back to the definition of a homogeneous Markov
  chain on~$S^1$. The transition matrix $P_{i,j}$ would then be given
  by $P_{i,j}=\bpr_{(i,*)}\bigl(\uparrow(j,*)\bigr)$.
\item Contrast this definition with an alternative, naive model
  consisting of a Markov chain on the state of global states. Note
  that the Markov property stated in Eq.~\eqref{eq:4} is relative to
  any ``cut'' $\gamma(s)$ of the trajectory. However, for a Markov
  chain, the property would only hold for particular cuts, namely
  those such that $|s|$ has the form $(n,n)$ for some integer~$n$.
\end{enumerate}

\noindent Checking that a probabilistic process $\pr$ satisfies the Markov
  property amounts to verifying the equality: 
\begin{equation}
\label{eq:7}
\frac1{\bpr_\alpha(s)}{\bpr_\alpha(s\cdot t)}=\bpr_{\gamma(s)}(t)
\end{equation}
for all finite trajectories $s$ and $t$ such that $\bpr_\alpha(s)>0$.
The following lemma however shows that, for closed processes, it
suffices to verify Eq.~\eqref{eq:7} for elementary trajectories~$t$.

\begin{lem}
  \label{lem:1}
  Let\/ $\pr=(\bpr_\alpha)_{\alpha\in X_0}$ be a closed \tcp, such that:
  \begin{equation}
    \label{eq:10}
    \forall\alpha\in X_0\qquad
   \bigl(\bpr_\alpha\bigr)_s(t)=\bpr_{\gamma(s)}(t),
  \end{equation}
  for every elementary trajectory $t$ and finite trajectory $s$ with\/
  $\bpr_\alpha(s)>0$. Then\/ $\pr$ is a Markov \tcp.
\end{lem}

\proof
  Let $\EE$ denote the set of elementary trajectories
  (Definition~\ref{def:5}). We also denote by $\EE^+$ the set of trajectories
  that are finite concatenations of elementary trajectories, and by
  $\VV$ the set of finite trajectories.  We proceed in two steps to
  show that Eq.~\eqref{eq:10} is valid for $s,t\in\VV$.

\par\medskip
\textsl{Step~1: Equation~\eqref{eq:10} is true for $s\in\VV$ and
  $t\in\EE^+$.}\quad By induction, we show that Eq.~\eqref{eq:10} is
true for $s\in\VV$ and $t=t_1\cdot\ldots\cdot t_n$ with
$t_i\in\EE$\,. The case $n=1$ is given by the hypothesis of the lemma,
assume it is true for all $k<n$. Assume moreover that
$\bpr_\alpha(s\cdot t_1\cdot\ldots\cdot t_k)>0$ for all
$k=1,\ldots,n-1$.  We calculate as follows, using the
hypothesis of the lemma and the induction hypothesis:
    \begin{align}
\notag
      \bigl(\bpr_\alpha\bigr)_s(t_1\cdot\ldots\cdot t_n)&=
\frac{\bpr_\alpha(s\cdot t_1\cdot\ldots\cdot t_n)}
{\bpr_\alpha(s)}\\
\notag
&=\bigl(\bpr_\alpha\bigr)_s(t_1\cdot\ldots\cdot t_{n-1})\cdot
\bigl(\bpr_\alpha\bigr)_{s\cdot t_1\cdot\ldots\cdot t_{n-1}}(t_n)\\
\label{eq:38}&=\bpr_{\gamma(s)}(t_1\cdot\ldots\cdot t_{n-1})\cdot
\bpr_{\gamma(t_{n-1})}(t_n).
    \end{align}

    We also have, using again the hypothesis of the lemma:
\begin{align}
\notag
\bpr_{\gamma(s)}(t_1\cdot\ldots\cdot
t_n)&=\bpr_{\gamma(s)}(t_1\cdot\ldots\cdot
t_{n-1})\cdot\bpr_{\gamma(s)}(t_1\cdot\ldots\cdot t_n|t_1\cdot\ldots\cdot
t_{n-1})\\
\notag
&=\bpr_{\gamma(s)}(t_1\cdot\ldots\cdot
t_{n-1})\cdot \bigl(\bpr_{\gamma(s)}\bigr)_{t_1\cdot\ldots\cdot
  t_{n-1}}(t_n)\\
\label{eq:32}
&= \bpr_{\gamma(s)}(t_1\cdot\ldots\cdot
t_{n-1})\cdot\bpr_{\gamma(t_{n-1})}(t_n)\,.
\end{align}
Comparing \eqref{eq:38} and \eqref{eq:32}, we get
$
 \bigl(\bpr_\alpha\bigr)_s(t_1\cdot\ldots\cdot t_n)=
\bpr_{\gamma(s)}(t_1\cdot\ldots\cdot t_n)$\,,
completing the induction in this case.

To be complete, we examine the case where $\bpr_\alpha(s\cdot
t_1\cdot\ldots\cdot t_k)=0$ for some integer
$k\in\{1,\ldots,n-1\}$. Then, on the one hand, this implies
$\bpr_\alpha(s\cdot t_1\cdot\ldots\cdot t_n)=0$ and thus
$\bigl(\bpr_\alpha\bigr)_s(t_1\cdot\ldots\cdot t_n)=0$. On the other
hand, let $i$ be the smallest integer $1\leq i<n$ such that
$\bpr_\alpha(s\cdot t_1\cdot\ldots\cdot t_i)=0$. Then the minimality
of $i$ yields $\bigl(\bpr_\alpha\bigr)_{s\cdot t_1\cdot\ldots\cdot
  t_{i-1}}(t_i)=0$, and by the hypothesis of the lemma this is
$\bpr_{\gamma(t_{i-1})}(t_i)=0$\,. Applying again the hypothesis of
the lemma: $\bpr_{\gamma(s)}(t_1\cdot\ldots\cdot
t_i)=\bpr_{\gamma(s)}(t_1\cdot\ldots\cdot
t_{i-1})\cdot\bpr_{\gamma(t_{i-1})}(t_i)=0$, which implies
$\bpr_{\gamma(s)}(t_1\cdot\ldots t_n)=0$. The induction is complete.

\par\medskip
\textsl{Step~2: Equation~\eqref{eq:10} is true for $s,t\in\VV$.}\quad
Let $s$ and $t$ be any finite trajectories. For
$\omega\in\,\uparrow(s\cdot t)$, we put
\begin{equation}
    \label{eq:12}
    \notag
E_\omega=\{v\in\EE^+\tq
s\cdot t\leq v\leq\omega\}\,,\qquad
\omega_T=\inf E_\omega\,.
\end{equation}
On the one hand, $E_\omega\neq\emptyset$ $\bpr_\alpha$-\as\ since $\pr$ is
assumed to be closed. On the other hand, the trajectories of $E_\omega$
form a chain, which is well founded by Lemma~\ref{lem:5}; hence
$\omega_T=\min E_\omega$ is $\bpr_\alpha$-\as\ well defined and
$\omega_T\in E_\omega$\,. It is easy to observe that, for $v=\omega_T$\,, we
have:
\begin{equation}
  \label{eq:13}
\{\omega'\in\Omega\tq\omega'_T=v\}=\uparrow v.
\end{equation}
(Later, we will interpret this by saying that $\omega\mapsto\omega_T$
is a stopping time). Since $\omega_T$ ranges over finite trajectories,
the set of values it can take is countable. Therefore,
decomposing with respect to the possible values:
\begin{align*}
  \label{eq:14}
  \notag
  \bigl(\bpr_\alpha\bigr)_s(t)&=\sum_v\bigl(\bpr_\alpha\bigr)_s(\uparrow
  t\cap\{\omega_T=v\})\\
&=\sum_{v}\bigl(\bpr_\alpha\bigr)_s(\omega_T=v)&&\text{since $\uparrow
  t\subset\{\omega_T=v\}$}\\
&=\sum_{v}\bigl(\bpr_\alpha\bigr)_s(\uparrow v)&&\text{by
  Eq.~\eqref{eq:13}}\\
&=\sum_{v}\bpr_{\gamma(s)}(\uparrow v)&&\text{by
Step~1 since $v\in\EE^+$}\\
&=\bpr_{\gamma(s)}(t)&&\text{recomposing.}
\end{align*}
The proof is complete.\qed

\section{Synchronization of Two Markov Chains}
\label{sec:synchr-two-mark}

In this section we introduce a way of constructing \abtcp s. It first
shows that our object of study is not empty. It also provides a bridge
between \abtcp s and usual Markov chains---another, maybe deeper link
is developed in~\S~\ref{sec:stopp-times-strong}.

Consider two Markov chains $\seq{X^1}n$ and $\seq {X^2}n$ on $S^1$ and
$S^2$ respectively. We denote by $M^i_x$ the probability measure on
$\Omega^i$ associated with the chain $X^i$ starting from state $x\in
S^i$, for $i=1,2$.  We assume for simplicity that both transition
matrices have all their coefficients positive. The construction consists of
recursively forcing the next synchronization of the chains on a shared
state. The formal construction is given in Definition~\ref{def:4} below,
after an informal explanation. The case where there is only one
synchronization state is trivial, in the sense that it reduces to the
independent product of the two Markov chains as shown by
Proposition~\ref{prop:8}, point~\ref{item:5}. A numerical example with two
synchronization states is analyzed in~\S~\ref{sec:few-examples}.

Denoting as above by $Q$ the set $S^1\cap S^2$ of shared states, let
$\tau^i$ be the first hitting time of $Q$ for the chain~$X^i$, defined
on $\Omega^i$ by
\begin{equation}
  \label{eq:5}
  \notag
\tau^i=\inf\{n>0 \tq X^i_n\in Q\},\quad \text{noting that
  $\tau^i<\infty$ $M^i_x$-a.s.}
\end{equation}
We consider  the subset  $X_0$ of global states given by
\begin{equation}
\notag  \label{eq:6}
X_0=\{(x,z)\in S^1\times S^2\tq x\in Q\wedge z\in Q\Rightarrow x=z\}.
\end{equation}
Introduce also $\Delta=\bigl\{(\tau^1<\infty)\wedge(\tau^2<\infty)
\wedge(X^1_{\tau^1}=X^2_{\tau^2})\bigr\}$, a measurable subset of
$\Omega^1\times\Omega^2$. Since the transition matrices we consider
have all their coefficients positive, we have $M_x^1\otimes
M_z^2(\Delta)>0$ for any global state~$(x,z)$. We therefore equip the
random pair of sequences $\sigma_0=\bigl(X^1_1X^1_2\dots
X^1_{\tau^1},X^2_1X^2_2\dots X^2_{\tau^2}\bigr)$ with the conditional
law 
\begin{equation}
\notag\label{eq:9}
U_{(x,z)}(\,\cdot\,)=M^1_x\otimes M^2_z(\,\cdot\;|\Delta).
\end{equation}
Starting now from the global state $(X^1_{\tau^1},X^2_{\tau^2})$, we
consider a fresh copy $\sigma_1$ of the same random pair of sequences,
now equipped with the law $U_{(X^1_{\tau^1},X^2_{\tau^2})}$\, (observe
that, by construction, $X^1_{\tau^1}=X^2_{\tau^2}$).

We construct inductively in this way a sequence $\seq\sigma n$ of
random trajectories, for which the concatenation
$\omega=\sigma_0\cdot\sigma_1\cdot\ldots$ is an element of~$\Omega$
since $|\sigma_k|\geq(1,1)$ for all $k\geq0$. Denoting by
$\bpr_{(x,z)}$ the law of $\omega$ thus constructed, we obtain a
probabilistic \tcp\ (Definition~\ref{def:1}), which is a closed process by
construction. 

\medskip

\begin{defi}
  \label{def:4}
The \tdefine{synchronization of the two Markov chains} $\seq {X^1}n$
and $\seq{X^2}n$ is the probabilistic process
$\pr=(\bpr_\alpha)_{\alpha\in X_0}$, where:
\begin{enumerate}[(1)]
\item $\alpha$ ranges over the set $X_0=\{(x,z)\in S^1\times
  S^2\tq x\in Q\wedge z\in Q\Rightarrow x=z\}$\,.
\item $\bpr_{(x,z)}$ is defined as the law of the infinite
  concatenation $\sigma_1\cdot\sigma_2\cdot\ldots$, where $\seq\sigma
  n$ is the countable Markov chain on the set $\EE$ of elementary
  trajectories with $U_{(x,z)}=M^1_x\otimes M^2_z(\,\cdot\;|\Delta)$
  as initial law, and transition kernel $K$ given by:
\begin{equation}
  \label{eq:8}
  \notag
\forall\sigma,\sigma'\in\EE,\qquad 
K(\sigma,\sigma')=U_{\gamma(\sigma)}(\sigma').
\end{equation}
\end{enumerate}\medskip
\end{defi}

\noindent Translating the above definition in the two-components processes
language consists of determining the value of $\bpr_\alpha(\uparrow
v)$ for any finite trajectory~$v$. This can be easily done only for
$v$ of the following special form:
\begin{equation*}
  v=\sigma_1\cdot\ldots\cdot\sigma_n\,,\quad\sigma_i\in\EE,
\qquad  \bpr_\alpha(\uparrow v)=
  U_\alpha(\sigma_1)\cdot K(\sigma_1,\sigma_2)\cdot\ldots
\cdot K(\sigma_{n-1},\sigma_n)\,.
\end{equation*}
Note that this entirely determines the probability~$\bpr_\alpha$\,;
since $\bpr_\alpha(v)$ for any finite trajectory $v$ will be computed
as the sum of all $\bpr_\alpha(w)$, for $w$ of the form
$w=\sigma_1\cdot\ldots\cdot\sigma_n$ and $v\leq w$, very much as we
did in Step~$2$ in the proof of Lemma~\ref{lem:1}.

\medskip

\begin{thm}
  \label{thr:1}
The synchronization of two Markov chains is a Markov \tcp.
\end{thm}

\proof Let $\alpha=(x,z)$ be an initial state.  Let $t\in\EE$ be any
elementary trajectory, and let $s$ be any finite trajectory. Denote
the coordinates of trajectories on each site by \mbox{$s=(s^1,s^2)$}
and \mbox{$t=(t^1,t^2)$}, and put $\gamma(s)=(x',z')$. Applying
Lemma~\ref{lem:1}, we have to show that
\mbox{$\bigl(\bpr_\alpha\bigr)_s(t)=\bpr_{\gamma(s)}(t)$}. We proceed
in two steps.
  \begin{enumerate}[(1)]
  \item \textsl{Step~$1$: $s$ is synchronization free.} Then $s\cdot
    t$ is an elementary trajectory, and by construction of $\sigma_0$
    we have $\uparrow(s\cdot t)=\{\sigma_0=s\cdot t\}$ and thus
    $\bpr_\alpha(s\cdot t)=U_\alpha(s\cdot t)$ by construction. From
    this we compute:
  \begin{equation}
\label{eq:15}
\bigl(\bpr_\alpha\bigr)_s(t)=
\frac{M_x\otimes M_z\bigl(\uparrow(s\cdot t)\cap\Delta\bigr)}
{M_x\otimes M_z(\uparrow s\cap\Delta)}\,.
\end{equation}
On the one hand, noting that $\uparrow(s\cdot t)\subset\Delta$, we
  have $M_x\otimes M_z\bigl(\uparrow(s\cdot
  t)\cap\Delta\bigr)=M_x(s^1\cdot t^1)M_z(s^2\cdot t^2)$. On the other
  hand, we have $M_x\otimes M_z(\uparrow s\cap\Delta)=M_x(s^1)M_z(s^2)
  M_x\otimes M_z(\Delta|\uparrow s)$ and, since $M_x$ and $M_z$ are
  Markov chains, $M_x\otimes M_z(\Delta|\uparrow s)=M_{x'}\otimes
  M_{z'}(\Delta)$. Going back to Eq.~\eqref{eq:15} we get:
  \begin{align*}
    \bigl(\bpr_\alpha\bigr)_s(t)
	&=\frac{M_x(s^1\cdot t^1)}{M_x(s^1)}\times
        \frac{M_z(s^2\cdot t^2)}{M_z(s^2)}\times
        \frac1{M_{x'}\otimes
          M_{z'}(\Delta)}\\
	&=M_{x'}(t^1)M_{z'}(t^2)\frac1{M_{x'}\otimes
          M_{z'}(\Delta)}\\
	&=M_{x'}\otimes M_{z'}(t|\Delta)=\bpr_{\gamma(s)}(t).
  \end{align*}

\item \textsl{Step $2$: $s$ is any finite trajectory.} Let
  $s=\sigma_0\cdot\sigma_1\cdot\ldots\cdot\sigma_p\cdot s'$ be the
  decomposition of $s$ according to Proposition~\ref{prop:1},
  case~\ref{item:10}, so that $\sigma_0,\dots,\sigma_p$ are elementary
  trajectories, and $s'$ is a synchronization free trajectory (the
  case $s'=\emptyset$ is admissible). We compute:
  \begin{align*}
    \bigl(\bpr_\alpha\bigr)_s(t)&=
\frac{\bpr_\alpha(\sigma_0\cdot\ldots\cdot\sigma_p\cdot s'\cdot t)}
{\bpr_\alpha(\sigma_0\cdot\ldots\cdot\sigma_p\cdot s')}  \\
&=\frac{U_{(x,z)}(\sigma_0)K(\sigma_0,\sigma_1)\ldots
  K(\sigma_{p-1},\sigma_p)  K(\sigma_p,s'\cdot t)}
{U_{(x,z)}(\sigma_0)K(\sigma_0,\sigma_1)\ldots
  K(\sigma_{p-1},\sigma_p) U_{\gamma(\sigma_p)}(\uparrow s')}\\
&=U_{\gamma(\sigma_p)}(s'\cdot t|\uparrow
s')\\
&=\bigl(\bpr_{\gamma(\sigma_p)}\bigr)_{s'}( t)\\
&=\bpr_{\gamma(s)}(t),
  \end{align*}
  the last equality following from Step~$1$ together with
  $\gamma(s')=\gamma(s)$.  
\end{enumerate}

Conclusion: Lemma~\ref{lem:1} applies, and
  $(\bpr_\alpha)_{\alpha\in X_0}$ is a \abtcp.
\qed

\section{Stopping Times and the Asynchronous 
Strong Markov Property}
\label{sec:stopp-times-strong}

All the notions and results of this section do not depend on the
particular structure of trajectories, and in particular they do not
rest on Proposition~\ref{prop:1}. It follows that they have straightforward
generalizations to an asynchronous model with an arbitrary number
$n\geq2$ of sites.

\subsection{Stopping Times}
\label{sec:stopping-times}

Stopping times are a fundamental tool in the theory of probabilistic
processes in general, and in the theory of Markov chain in
particular. Recall that a stopping time associated to a Markov chain
is a random integer, maybe infinite and seen as a random time instant,
with the following property: an observer aware of the successive values
of the chain can decide at each instant whether the stopping time has
already been reached or not. Standard examples of stopping times in
Markov chain theory are: constant times (trivial since non random);
the \emph{first} instant where the chain hits a given state; more
generally the first instant where a chain reaches a given set of
states. A standard example of a random time which is \emph{not} a
stopping time is the \emph{last} instant where the chain hits a given
state.

It is natural to introduce an equivalent notion for \tcp es, and this
is the topic of this subsection. We will see that the first instant a
process hits a global state defines a stopping time; but, and
contrasting with Markov chains, the first instant of reaching a given
\emph{set} of global states does not define a stopping time in
general, unless one considers special kinds of sets. 

Recall that $\TT=\bigl(\NN\times\NN\bigr)\cup\{\infty\}$ denotes the partially
ordered set of ``two-components time instants''.

\begin{defi}[Random times and stopping times]
\label{def:6}
Let $T:\Omega\to\TT$ be an arbitrary mapping. For any
$\omega\in\Omega$, we denote by $\omega_T$ the prefix of $\omega$ of
length $T(\omega)$ if $T(\omega)<\infty$, and we put $\omega_T=\omega$
if $T(\omega)=\infty$.
\begin{enumerate}[(1)]
\item We say that $T$ is a \tdefine{random time} if $\omega_T$ is a
  subtrajectory of $\omega$ for all $\omega\in\Omega$.
\item If $T$ is a random time we say that $T$ is a \tdefine{stopping
    time} if furthermore the following property holds:
\begin{equation}
\label{eq:45}
  \forall\omega,\omega'\in\Omega\quad 
\omega'\geq \omega_T\Rightarrow 
\omega_T=\omega'_T\,.
\end{equation}
\end{enumerate}
\end{defi}

\noindent Actually since the space $\Omega$ is always implicitly equipped with
an initial state~$\alpha$, a more general notion of stopping times
would be as for probabilistic processes a \emph{family} of random
times~$(T_\alpha)_{\alpha\in X_0}$\,, each one satisfying
condition~(\ref{eq:45}). But, since we will only be concerned with
stopping times independent of~$\alpha$, we prefer limiting ourselves
to Definition~\ref{def:6} as it is formulated.

\medskip Since $\TT$ is a countable set, it is naturally equipped with
its discrete \slgb. It turns out that a stopping time
$T:\Omega\to\TT$ is always measurable; and so is the mapping
$\omega\in\Omega\mapsto\omega_T$\,, provided we equip the set of
trajectories (either finite or infinite) with the \slgb\ generated by
the sets of the form $\{v\tq s\leq v\}$, for $s$ ranging over finite
trajectories, and $v$ ranging over trajectories. If the set of
trajectories is seen as a DCPO (Directed Complete Partial
Order~\cite{gierz03}), this is the Borel \slgb\ associated with the
Scott topology on the DCPO.  Obviously, it induces by restriction the
\slgb\ $\FFF$ on the subset~$\Omega$.

\begin{prop}
  \label{prop:4}
  Let $T:\Omega\to\TT$ be a stopping time. We equip $\TT$ with its
  discrete $\slgb$, and we equip the set of trajectories with its
  Borel \slgb\ described above.
  \begin{enumerate}[\em(1)]
  \item Then $T$ and $\omega_T$ are two measurable mappings.
  \item Let $\FFF_T$ denote the \slgb\ generated by~$\omega_T$\,. Then
    $\FFF_T$ is finer than the \slgb\ generated by~$T$, and it is 
    characterized as follows:
  \begin{equation*}
    \forall A\in \FFF\quad A\in\FFF_T\iff
    \forall\omega,\omega'\in\Omega\quad \omega\in
    A\wedge\omega'\geq\omega_T\Rightarrow\omega'\in A.
  \end{equation*}
  \end{enumerate}
\end{prop}

\proof
If $Y:(\Omega,\FFF)\to (A,\GGG)$ is a measurable mapping, we denote by
$\langle Y\rangle$ the sub-\slgb\ of $\FFF$ generated by~$Y$, and given
by $\langle Y\rangle =\{Y^{-1}(U)\tq U\in\GGG\}$. 
  For $\omega\in\Omega$, let $\zeta(\omega)=\omega_T$.  For any
  finite trajectory~$v$, we put
$$
S_v=\{\text{$w$ trajectory}\tq v\leq w\}\,.
$$
Since $T$ is a stopping time, $\zeta^{-1}\bigl(\{v\}\bigr)$~is either
empty or equal to~$\uparrow v$, so $\zeta^{-1}\bigl(\{v\}\bigr)$ is
measurable in either cases. Let us denote by $\VV$ the set of finite
trajectories. Since $\VV$ is countable, it follows that
$\zeta^{-1}(S_v\cap\VV)=\bigcup_{w\in
  S_v\cap\VV}\zeta^{-1}\bigl(\{w\}\bigr)$ is measurable for any
$v\in\VV$, as well as $\zeta^{-1}(\VV)$. By definition of
$\zeta=\omega_T$, we have that $\zeta(\omega)$ is either finite or
maximal. From this, it follows first that
$\zeta^{-1}(\Omega)=\Omega\setminus\zeta^{-1}(\VV)$ is measurable; and
second:
\begin{equation*}
\zeta^{-1}(S_v)=\zeta^{-1}(S_v\cap\VV)\cup\zeta^{-1}(\uparrow v).
\end{equation*}
But $\zeta^{-1}(\uparrow v)=\zeta^{-1}(\Omega)\cap\uparrow v$, hence
$\zeta^{-1}(S_v)$ is the union of two measurable subsets of~$\Omega$,
and is thus measurable. This shows that $\zeta$ is a measurable
mapping. 

To prove that $T$ is measurable, observe that:
\begin{equation}
    \label{eq:20}
    \notag
\forall (m,n)\in\TT\quad
\bigl\{T=(m,n)\bigr\}=\bigcup_{|v|=(m,n)}\zeta^{-1}(v)\,.
\end{equation}
Since the union is finite, it follows that $\{T=(m,n)\}$ is a
$\langle\zeta\rangle$-measurable subset, from which we deduce that
$\{T=\infty\}=\bigcup_{(m,n)\in\NN\times\NN}\{T\neq(m,n)\}$ is also a
$\langle\zeta\rangle$-measurable subset. Therefore $\langle
T\rangle\subset\langle\zeta\rangle$. By the property of stopping times
$\omega'\geq\omega_T$ is equivalent to $\omega_T=\omega'_T$\,, from
which follows the characterization of $\FFF_T=\langle\zeta\rangle$.
\qed

Note that any function $f:(\Omega,\FFF)\to(A,\GGG)$ with value in
some measurable space is measurable with respect to $\FFF_T$ if and
only if it is constant on elementary cylinders of the form $\uparrow
v=\{\omega_T=v\}$ with  $v$ ranging over the values of~$\omega_T$---since
it is well known that $f$ is $\FFF_T$-measurable if and only if it
can be written as $f(\omega)=g(\omega_T)$ with $g$ some measurable
mapping.  

\subsection{Shift Operators}

In Markov chain theory, the ``universal'' shift operator $\theta$ is
classically defined on the space of trajectories of a Markov chain by
$\theta(x_0x_1\ldots)=(x_1x_2\ldots)$. Its iterations $\theta_n$ are
defined for $n\geq0$ by $\theta_0=\text{Id}$ and
$\theta_{n+1}=\theta\circ\theta_n$\,. Allowing the time index $n$ to
be random, one defines~$\theta_T$, for $T:\Omega\to \NN$ any random
variable, by $\theta_T(\omega)=\theta_{T(\omega)}(\omega)$. In our
framework, there is no such ``universal'' shift
operator~$\theta$. Yet, each stopping time $T:\Omega\to\TT$ induces a
shift operator~$\theta_T:\Omega\to\Omega$. Informally
$\theta_T(\omega)$ is the queue of trajectory $\omega$ that remains
``after'' the prefix trajectory~$\omega_T$\,.

\begin{defi}
\label{def:7}
Let $T:\Omega\to\TT$ be a stopping time.  The \tdefine{shift operator}
associated with $T$ is the mapping $\theta_T:\Omega\to\Omega$ , which
is only partially defined; if $T(\omega)<\infty$, then
$\theta_T(\omega)$ is defined as the unique element of\/ $\Omega$ such
that
\begin{equation*}
\omega=\omega_T\cdot \theta_T(\omega),
\end{equation*}
and $\theta_T(\omega)$ is undefined otherwise.
\end{defi}

The shift operator allows to define an  addition on stopping times, as
shown by the following result which mimics an equivalent result widely
used in Markov chain theory. 

\begin{lem}
  \label{lem:8}
Let $S,T:\Omega\to\TT$ be two stopping times. Then $U=S+T\circ\theta_S$
is a stopping time (it is understood that $U=\infty$ if $S=\infty$).
\end{lem}

\proof
  It is clear that $U$ is a random time.  Let
  $\omega,\omega'\in\Omega$ such that $\omega'\geq\omega_U$, we have
  to show that $\omega'_U=\omega_U$. If $S(\omega)=\infty$ then
  $U(\omega)=\infty$ and then $\omega'=\omega$ and
  $\omega_U=\omega'_U$, trivially.

  Hence we assume without loss of generality that $S(\omega)<\infty$,
  and we put $\zeta=\theta_S(\omega)$ and $\zeta'=\theta_S(\omega')$.
  We have $\omega_U=\omega_S\cdot \zeta_T$. Hence
  $\omega'\geq\omega_S$ and thus $\omega'_S=\omega_S$ since $S$ is a
  stopping time. Therefore:
  $\omega'=\omega_S\cdot\zeta'\geq\omega_U=\omega_S\cdot \zeta_T$\,,
  hence $\zeta'\geq\zeta_T$\,. Thus $\zeta'_T=\zeta_T$ since $T$ is a
  stopping time.  We have finally
  $\omega'_U=\omega'_S\cdot\zeta'_T=\omega_S\cdot\zeta_T=\omega_U$,
  proving that $U$ is a stopping time.
\qed

Starting from a given stopping time~$T$, we use Lemma~\ref{lem:8} above
 to iterate the ``addition''  of~$T$ to itself. 

\begin{defi}
\label{def:15}
  Let $T:\Omega\to\TT$ be a stopping time, and let $\theta_T$ be the
  associated shift operator. The sequence $(T^n)_{n\geq0}$ of
  mappings $\Omega\to\TT$ defined as follows:
\begin{align*}
T^0&=(0,0)&
\forall n\geq0\quad T^{n+1}&=T^n+T\circ\theta_{T^n}
\end{align*}
with the convention that $T^{n+1}=\infty$ on $\{T^n=\infty\}$, 
is a sequence of stopping times, called \tdefine{iterated stopping
  times associated with~$T$}.
\end{defi}

Remark that  $\theta_{T^0}=\text{Id}_\Omega$, and $T^1=T$.

\subsection{Examples of Stopping Times}
\label{sec:exampl-stopp-times}

In this subsection we review some examples of random times, and
analyze whether they are stopping times or not. Some of the examples
introduced here will be used later
in~\S\S~\ref{sec:recurr-transc-glob}--\ref{sec:open-closed-tcp}. 

\subsubsection{Constant Times are not Random Times in General.}
In general, if $(m,n)\in\NN\times\NN$, then the random variable
constant and equal to $(m,n)$ is not a random time. For instance, take
$(m,n)=(2,2)$ and consider as in the Introduction a maximal trajectory
$\omega$ starting with $(a\cdot\c,e\cdot f\cdot\c)$ with $\c$ as
synchronization state. Then the prefix of length $(2,2)$ of $\omega$
is $(a\cdot\c,e\cdot f)$, which is not a trajectory. Hence the
constant $T=(2,2)$ is not a random time. This contrasts with Markov
chain theory, where constant times are a basic example of stopping
times.

However note that any constant time is indeed a random time if the
process is open (Definition~\ref{def:17}). And in this case, it is also a
stopping time.

\subsubsection{A Random Time which is not a Stopping Time.}
For $\omega$ a maximal trajectory, let $v$ be the
first elementary trajectory in the decomposition of $\omega$ as in
Proposition~\ref{prop:1}, which is defined if $\omega$ has at least one
synchronization. Then $v$ has the form $v=u\cdot(y,y)$ for some unique
finite trajectory $u$ and state $y\in Q$. Put $\omega_T=u$ in this
case, and $\omega_T=\omega$ if $v$ is not defined. Time $T(\omega)$
represents the ``last instant before first synchronization''. By
construction, $T$~is a random time since $\omega_T$~is a subtrajectory
of~$\omega$. 

However $T$ is not a stopping time in general. For example, consider
$S^1=\{a,b,\c,\d\}$ and $S^2=\{\c,\d,e,f\}$, if $\omega$ starts with
$(a\cdot\c,f\cdot f\cdot e\cdot\c)$, then $T(\omega)=(1,3)$ and
$\omega_T=(a,f\cdot f\cdot e)$, corresponding to the last private
states $a$ and $e$ before synchronization on $(\c,\c)$. And if
$\omega'$ starts with $(a\cdot b\cdot\c,f\cdot f\cdot e\cdot
f\cdot\c)$, then $T(\omega')=(2,3)\neq T(\omega)$ although
$\omega'\geq \omega_T$\,. This shows that $T$ is not a stopping time.

\subsubsection{The First Return Time of a Global State.}
\label{sec:first-hitting-time}

Let $\alpha\in X_0$ be a given global state. For
$\omega\in\Omega$, consider the following set of finite
subtrajectories of~$\omega$:
\begin{equation*}
 N_\alpha(\omega)=\{v\leq\omega\tq
\text{$v$ finite subtrajectory of $\omega$}
\wedge 
\gamma(v)=\alpha\wedge |v|\geq(1,1)\}.
\end{equation*}
If nonempty, $N_\alpha$ is a sublattice of the lattice of
subtrajectories of~$\omega$ since, by Lemma~\ref{lem:5}, lower bounds
are taken component by component so that $\gamma(v\wedge v')=\alpha$
whenever $v,v'\in N_\alpha(\omega)$ and $|v\wedge v'|\geq(1,1)$. In
particular, if we put $v=\min N_\alpha(\omega)$, which exists whenever
\mbox{$N_\alpha(\omega)\neq\emptyset$}, then $\gamma(v)=\alpha$ and
$|v|\geq(1,1)$. We define thus the first return time to $\alpha$ as
follows.

\begin{defi}
\label{def:8}
  For any $\alpha\in X_0$\,, the \tdefine{first return time
    to~$\bm\alpha$} is 
the stopping time $R_\alpha:\Omega\to\TT$ defined by:
\begin{equation*}
\forall\omega\in\Omega\quad   \omega_{R_\alpha}=
  \begin{cases}
    \omega,&\text{if $N_\alpha(\omega)=\emptyset$,
and thus $R_\alpha(\omega)=\infty$,}\\
    \min N_\alpha(\omega),&\text{otherwise, and thus
      $R_\alpha(\omega)=\bigl|\min N_\alpha(\omega)\bigr|$.}
  \end{cases}
\end{equation*}
The \tdefine{successive return times to~$\bm\alpha$} are the iterated
stopping times $(R^n_\alpha)_{n\geq1}$ associated with~$R_\alpha$ as
in Definition~{\normalfont\ref{def:15}}.
\end{defi}

For any finite subtrajectory $v$ of~$\omega$, we have:
\[
\bigl(\gamma(v)=\alpha\wedge |v|\geq(1,1)\bigr)\Longrightarrow
\omega_{R_\alpha}\leq v\,,
\]
which is consistent with the intuition of what a ``first return time''
should be. To show that $R_\alpha$ is indeed a stopping time, observe
first that $\omega_{R_\alpha}$ is clearly a subtrajectory
of~$\omega$. And second, if $\omega'\in\Omega$ is such that
$\omega'\geq\omega_{R_\alpha}$, that implies that
$\omega_{R_\alpha}\in N_\alpha(\omega')$, and thus
$\omega'_{R_\alpha}\leq\omega_{R_\alpha}$ by minimality
of~$\omega'_{R_\alpha}$\,. But then $\omega'_{R_\alpha}\in
N_\alpha(\omega)$, and thus $\omega_{R_\alpha}\leq\omega'_{R_\alpha}$
by minimality of~$\omega_{R_\alpha}$\,. Hence
$\omega_{R_\alpha}=\omega'_{R_\alpha}$\,, and this shows that
$R_\alpha$ is a stopping time.

\begin{figure}
\setbox1=\hbox{$\bullet$}
\newcommand{\dbullet}{\rlap{\raisebox{-2.5 pt}{$\bullet$}}\raisebox{2.5 pt}{$\bullet$}}
\newcommand{\lu}[2]{\POS[]!U(#1)\drop{#2}}
\newcommand{\ld}[2]{\POS[]!D(#1)\drop{#2}}
\renewcommand{\ll}[1]{\POS[]!L(6)\drop{#1}}
  \centering
  \[
\xymatrix@M=0pt{
    \bullet\ar@{-}[rr]!R\lu5{a}
  &&\bullet\ar@{-}[dr]\lu5{b}
  &&\ar@{-}[r]
   &\bullet\ar@{-}[r]\lu5{a}
   &\ar@{-}[dr]&&\\
 &&&\dbullet\ar@{-}[ur]\ar@{-}[dr]\lu3{\c}\ld{2.666}{\c}
&&&&\dbullet\ar@{-}[ur]\ar@{-}[dr]\lu3{\d}\ld{2.666}{\d}&\\
   &\bullet\ar@{-}[r]!R\ld4e
   &\ar@{-}[ur]
  &&\bullet\ar@{-}[rr]\ld4e
  &&\bullet\ar@{-}[ur]\ld4f&&
}
\]
  \caption{\textsl{A finite trajectory synchronizing
      on shared states $\c$ and~$\d$.}}
  \label{fig:mkjsdfqqs}
\end{figure}
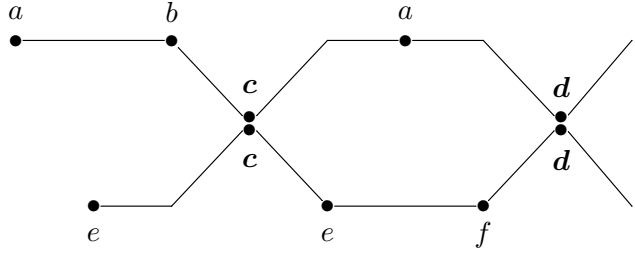

\medskip As an example, consider $S^1=\{a,b,\c,\d\}$ and
$S^2=\{\c,\d,e,f\}$, and a maximal trajectory $\omega$ starting with
\mbox{$(a\cdot b\cdot\c\cdot a\cdot \d,e\cdot\c\cdot e\cdot f\cdot\d)$},
which is depicted in Figure~\ref{fig:mkjsdfqqs}. Consider the global
state $\alpha=(a,e)$. Then $R_\alpha(\omega)=(1,1)$, and
$\omega_{R_\alpha}=(a,e)$. Note that, since $R_\alpha$ is indeed a
stopping time, we do not need to know the queue of $\omega$ to already
have information on $R(\omega)$.

Let us determine the value of next return $R^2_\alpha(\omega)$
to~$\alpha$. The shifted trajectory $\theta_{R_\alpha}(\omega)$ starts
with $(b\cdot\c\cdot a\cdot\d,\c\cdot e\cdot f\cdot\d)$. Therefore
$R_\alpha\bigl(\theta_{R_\alpha}(\omega)\bigr)=(3,2)$, and
$R^2_\alpha(\omega)=(1,1)+(3,2)=(4,3)$.  Note that $R^3(\omega)$ is
undetermined at this stage.

If $\zeta$ is the trajectory \mbox{$(b\cdot b\cdot\ldots,e\cdot\cdot
  e\cdot \ldots)$}, with only $b$ on the first component and only $e$
on the second component, then $R_\alpha(\zeta)=\infty$ and
$\zeta_{R_\alpha}=\zeta$.

\subsubsection{Supremum of Stopping Times.}

If $S$ and $T$ are two stopping times, then the random time $S\vee T$
defined by $\omega_{S\vee T}=\omega_S\vee \omega_T$ is a stopping
time. For, if $\omega'\geq\omega_S\vee\omega_T$, then
$\omega'\geq\omega_S$ and $\omega'\geq\omega_T$, therefore
$\omega'_S=\omega_S$ and $\omega'_T=\omega_T$, hence $\omega'_{S\vee
  T}=\omega_{S\vee T}$\,. The same line of proof shows that the
supremum of any family of stopping times is a stopping time.

\subsubsection{The Infimum of Stopping Times may not be a Stopping Time.}

Contrasting with stopping times from Markov chain theory however, the
infimum of two stopping times $S$ and~$T$, defined by $\omega_{S\wedge
  T}=\omega_S\wedge\omega_T$\,, may not be a stopping time. Let us
consider an example. Let $S^1=\{a,b,\c\}$ and $S^2=\{\c,e,f\}$. Let
$\alpha=(a,e)$ and $\beta=(b,f)$, and let $S=T_\alpha$ and $T=T_\beta$
be the first return times to $\alpha$ and to $\beta$
respectively. Consider a trajectory $\omega$ starting with
\mbox{$(a\cdot b,f\cdot e)$}. Then \mbox{$\omega_S=(a,f\cdot e)$} and
\mbox{$\omega_T=(a\cdot b,f)$}, and thus $\omega_{S\wedge
  T}=(a,f)$. However, if $\omega'$ is the maximal trajectory defined
by \mbox{$\omega'=(a\cdot a\cdots,f\cdot f\cdots)$} we have
$\omega'\geq\omega_{S\wedge T}$ on the one hand, and
$\omega'_S=\omega'$ and $\omega'_T=\omega'$ on the other hand, so that
$\omega'_{S\wedge T}=\omega'\neq\omega_{S\wedge T}$. This show that
$S\wedge T$ is not a stopping time.

This example is specific to the asynchronous structure we consider,
since it makes use of the \emph{partially} ordered structure of
trajectories.

\subsubsection{First Return Time to a Square Set of Global States.}
\label{sec:first-hitting-time-1}

Since the infimum of stopping times is not a stopping time in general,
there is an issue for defining the first return time to a set of
global states. There is actually no obvious way of defining such a
thing in general, as the analysis of the above example reveals. 
The situation however becomes favorable if one considers a set of
states satisfying the following property.

\begin{defi}
  \label{def:9}
  We say a subset $A\subset X_0$ of global sets is a \tdefine{square
    set} if it has the form $A=X_0\cap(S'_1\times S'_2)$ where
  $S'_1\subset S_1$ and $S'_2\subset S_2$\,.
\end{defi}

A first example of a square set is $X_0$ itself. We will also
encounter the square set $(Q\times Q)\cap X_0$\,. If $\alpha=(x,z)$
and $\beta=(x',z')$, the smallest square set containing $\alpha$ and
$\beta$ is $\{\alpha,\beta,(x,z'),(x',z)\}$.

Assume that $A$ is a square set of global states. Define then, for any
$\omega\in\Omega$: 
\begin{align*}
N_A(\omega)&=\{v\leq\omega\tq
\text{$v$ finite subtrajectory of $\omega$}\wedge
\gamma(v)\in A\wedge|v|\geq(1,1)\}\,.
\end{align*}
Then $N_A(\omega)$ is a  sublattice of the lattice of finite
subtrajectories of $\omega$ whenever it is nonempty. Indeed, since
$A$ is a square set. The random time $R_A$ defined by
\begin{equation*}
  \omega_{R_A}=\min N_A(\omega)\,,
\end{equation*}
and by $R_A=\infty$ as usual when $N_A(\omega)$ is empty, is a
stopping time that satisfies $\gamma(\omega_{R_A})\in A$ whenever or
$R_A<\infty$. We define $R_A$ as the \emph{first return time} to the
square set~$A$. One furthermore checks that
$\omega_{R_A}=\bigwedge_{\alpha\in A}\omega_{R_\alpha}$\,, providing
an example of infimum of stopping times the result of which is indeed
a stopping time.

\medskip Let us examine the first return times associated with the
square sets $X_0$ and \mbox{$(Q\times Q)\cap X_0$}. In Markov chain
theory, $R_{X_0}$~would correspond to the constant time~$1$. But in
the asynchronous framework its action is less simple. Stopping time
$R_{X_0}$ can be described as follows: $\omega_{R_{X_0}}$~is the
smallest subtrajectory of $\omega$ with length $\geq(1,1)$.  In
particular, $R_{X_0}(\omega)$ is always \emph{finite}.

We detail the action of $R_{X_0}$ on an example. Consider
$S^1=\{a,b,\c,\d\}$ and $S^2=\{\c,\d,e,f\}$, and let $\omega$ be some
maximal trajectory starting with \mbox{$(a\cdot b\cdot\c\cdot a\cdot
  \d,e\cdot\c\cdot e\cdot f\cdot\d)$}, as depicted in
Figure~\ref{fig:mkjsdfqqs} above. The exercise consists in finding the
values of $R^n_{X_0}(\omega)$ for the first integers~$n$, where
$R^n_{X_0}$ denote the iterated stopping times associated with
$R_{X_0}$ as in Definition~\ref{def:15}.  Obviously
$R_{X_0}(\omega)=(1,1)$. The shifted trajectory
$\theta_{R_{X_0}}(\omega)$ starts with \mbox{$(b\cdot\c\cdot
  a\cdot\d,\c\cdot e\cdot f\cdot\d)$}. The smallest subtrajectory of
$\theta_{R_{X_0}}(\omega)$ of length at least $(1,1)$ is
\mbox{$(b\cdot\c,\c)$}, and thus
$R_{X_0}\bigl(\theta_{R_{X_0}}(\omega)\bigr)=(2,1)$. Hence
$R^2_{X_0}(\omega)=(1,1)+(2,1)=(3,2)$.  The finite trajectories
$$\omega_{R_{X_0}}=(a,e) \quad\text{and}\quad
\bigl(\theta_{R_{X_0}}(\omega)\bigr)_{R_{X_0}}=(b\cdot\c,\c)
$$ 
yield the following initial decomposition of~$\omega$:
$\omega=(a,e)\cdot(b\cdot\c,\c)\cdot \theta_{R^2_{X_0}} (\omega)$. For
the next values $n=3,4$ we find $R^3_{X_0}(\omega)=(4,3)$ and
$R^4_{X_0}(\omega)=(5,5)$, corresponding to the initial decomposition
$\omega=(a,e)\cdot(b\cdot\c,\c)\cdot(a,e)\cdot(\d,f\cdot\d)\cdots$.

\medskip Coming now to the square set $(Q\times Q)\cap X_0$\,, and
denoting by $R_Q$ the first return time associated with it, we may
rephrase the definition of infinite synchronization of trajectories
(Definition~\ref{def:17}) as follows: a maximal trajectory $\omega$
synchronizes infinitely often if
\mbox{$\omega\in\bigcap_{n\geq1}\{R_Q^n<\infty\}$}. A probabilistic process
$\pr$ is closed if $R_Q^n<\infty$ for all $n\geq1$ and
$\bpr_\alpha$-almost surely, for all $\alpha\in X_0$\,. It is open if
$R_Q=\infty$, $\bpr_\alpha$-almost surely and for all $\alpha\in
X_0$\,.

\medskip We end this series of examples with the following result which
will be useful in the study of recurrence of global states. It makes use
of the finitary assumption on the set of global states.

\begin{lem}
\label{lem:6}
Let $A$ be a square set. Denoting by $(R^n_A)_{n\geq1}$ the successive
returns to~$A$, \textsl{i.e.}, the iterated stopping times associated
with the first return time~$R_A$, and by $(R^n_\alpha)_{n\geq1}$ the
successive return times to~$\alpha$ for any $\alpha\in A$, we have the
following equality of sets:
\[
\bigcap_{n\geq1}\{R_A^n<\infty\}=
\bigcup_{\alpha\in A}\bigcap_{n\geq1}
\{R^n_\alpha<\infty\}\,.
\]
\end{lem}

\proof
  The $\supset$ inclusion is obvious. For the converse inclusion, let
  $\omega\in\Omega$ be such that $R^n_A(\omega)<\infty$ for all
  $n\geq1$. Since $A$ is a finite set, there exists some state
  $\alpha\in A$ and a strictly increasing sequence of integers
  $(n_k)_{k\geq1}$ such that $\gamma(R^{n_k}_A)=\alpha$ for
  all~$k$. By induction on~$k$, we show that $R^k_\alpha(\omega)\leq
  R_A^{n_k}(\omega)$ for all integers $k\geq1$. The finite trajectory
  $v=R_A(\omega)$ is a subtrajectory of $\omega$ satisfying
  $\gamma(v)=\alpha$ and $|v|\geq(1,1)$, and therefore
  $\omega_{R_\alpha}\leq v$. Since the sequence
  $\bigl(R^n_A(\omega)\bigr)_{n\geq1}$ is increasing, as shown by the
  formula in Definition~\ref{def:15} that defines it, we have
  $\omega_{R_\alpha}\leq
  v=\omega_{R^1_A}\leq\omega_{R^{n_1}_A}$\,. Assume for the induction
  that $R^k_\alpha(\omega)\leq R^{n_k}_A(\omega)$. Then there is some
  finite trajectory $v$ such that
  $\omega_{R^{n_k}_A}=\omega_{R^k_\alpha}\cdot v$. Since
  $n_k<n_{k+1}$, there is also some finite trajectory $v'$ such that
  $\gamma(v')=\alpha$, $|v'|\geq(1,1)$ and $\omega_{R_A^{n_{k+1}}}=
  \omega_{R_A^{n_{k}}}\cdot v'$\,. We obtain thus:
  \begin{equation*}
    \omega_{R^{n_{k+1}}_A}=\omega_{R^k_\alpha}\cdot v\cdot v',\quad
    |v\cdot v'|\geq(1,1),\quad\gamma(v\cdot v')=\alpha.
  \end{equation*}
  This implies that
  $R_\alpha\bigl(\theta_{R^k_\alpha}(\omega)\bigr)\leq|v\cdot v'|$. By
  definition, we have
  $R^{k+1}_\alpha=R^k_\alpha+R_\alpha\circ\theta_{R^k_\alpha}$,
  whence:
  \begin{equation*}
    \bigl|R^{k+1}_\alpha(\omega)\bigr|\leq
\bigl|R^{k}_\alpha(\omega)\bigr|+|v|+| v'|
= \bigl|R^{n_k}_A(\omega)\bigr|+| v'|=\bigl| R^{n_{k+1}}_A(\omega)\bigr|,
  \end{equation*}
  completing the induction. This implies in particular that
  $R^{k}_\alpha(\omega)<\infty$ for all $k\geq1$, as expected.
\qed

\subsection{The Asynchronous Strong Markov Property}
\label{sec:strong-mark-prop}

The Asynchronous Strong Markov Property that we state below has the
exact same formulation than the Strong Markov property for Markov
chains found in classical references~\cite[Theorem~3.5
p.23]{revuz75}. The syntactical identity underlines the parallel with
Markov chain theory, although the interpretation of symbols must be
changed of course: stopping times must be understood in the sense of
Definition~\ref{def:6}, the associated \slgb\ in the sense given in
Proposition~\ref{prop:4}, and of course \abtcp s replace Markov
chains. Nevertheless, once the Asynchronous Strong Markov property has
been established, it is possible to transfer verbatim some pieces of
Markov chain theory.  Examples of such transfers are Lemma~\ref{lem:7}
given just after Theorem~\ref{thr:2} and the \mbox{$0$-$1$} law for the
infinite return to a given global state, given in point~\ref{item:14}
of Proposition-definition~\ref{prop:3} below.

\medskip

\begin{thm}[Asynchronous Strong Markov property]
  \label{thr:2}
  Let\/ $\pr=(\bpr_\alpha)_{\alpha\in X_0}$ be a \abtcp. For any
  measurable and non negative function $h:\Omega\to\RRR$ and for any
  stopping time $T:\Omega\to\TT$, we have
\begin{equation}
  \label{eq:17}
  \forall\alpha\in X_0\qquad 
  \besp_\alpha(h\circ\theta_T\,|\,\FFF_T)=\besp_{\gamma(\omega_T)}(h)\,,\quad
\text{$\bpr_\alpha$-\as,}
\end{equation}
where $\besp_\alpha(\cdot|\FFF_T)$ denotes the conditional expectation
with respect to probability\/~$\bpr_\alpha$\, and
\slgb~$\FFF_T$\,. By convention, both sides of Eq.~\eqref{eq:17}
vanish outside $\{T<\infty\}$.
\end{thm}

Note that, as for the Strong Markov Property for Markov chains, both
sides of Eq.~\eqref{eq:17} are random variables: the left side, since
it is a conditional expectation with respect to \slgb~$\FFF_T$; and
the right side, since it depends on the random
variable~$\gamma(\omega_T)$.

\medskip

\proof
  Let $Z$ denote the random variable $Z=\besp_{\gamma(\omega_T)}(h)$,
  which is obviously $\FFF_T$-measurable since $\gamma(\omega_T)$
  is. Let $\phi$ be any non negative, bounded and $\FFF_T$-measurable
  function. Denote by $R$ the set of finite trajectories taken
  by~$\omega_T$.  Then, since $R$ is at most countable:
  \begin{equation}
\label{eq:22}
    \besp_\alpha\bigl(\phi\cdot
    h\circ\theta_T\bigr)=\sum_{\begin{substack}v\in R\end{substack}}
\besp_\alpha\bigl(\un{\omega_T=v}\cdot\phi\cdot h\circ\theta_T\bigr).
\end{equation}
Since $T$ is a stopping time, and since $T^{-1}(v)\neq\emptyset$ if
$v\in R$, we have $\{\omega_T=v\}=\uparrow v$. Furthermore, $\phi$~is
constant on $\{\omega_T=v\}$, so that if  $\phi(v)$ denote this
constant, we get:
\begin{align}
  \label{eq:23}
\notag
\besp_\alpha\bigl(\un{\omega_T= v}\cdot\phi\cdot h\circ\theta_T\bigr)&=
\phi(v)\bpr_\alpha\bigl(\uparrow v\bigr)
\frac{\besp_\alpha\bigl(\un{\uparrow v}h\circ\theta_T\bigr)}
{\bpr_\alpha\bigl(\uparrow v\bigr)}
\end{align}
Recognizing the conditional expectation defined as the future of $v$
w.t.r.\ to probability~$\bpr_\alpha$, we use the Markov
property~\eqref{eq:4} of Definition~\ref{def:3} to get:
\begin{equation}
  \label{eq:24}
  \notag
\besp_\alpha\bigl(\un{\omega_T= v}\cdot\phi\cdot h\circ\theta_T\bigr)=
\bpr_\alpha(\uparrow v)\phi(v)\besp_{\gamma(v)}(h).
\end{equation}
Going back to Eq.~\eqref{eq:22} we obtain:
\begin{align}
  \notag
\besp_\alpha\bigl(\phi\cdot
    h\circ\theta_T\bigr)&=\sum_{\begin{substack}v\in R\end{substack}}
\bpr_\alpha(\uparrow v)\phi(v)\besp_{\gamma(v)}(h)=\besp_\alpha
\bigl(\phi Z\bigr).
\end{align}
This shows that $Z=\besp_\alpha(h\circ\theta_T|\FFF_T)$.
\qed

The following result is a typical application of the Strong Markov
property in Markov chain theory that applies here too. It intuitively
says this: the probability of returning infinitely often to a
state~$\beta$, starting from~$\alpha$, is the product of the
probability of hitting $\beta$ once starting from~$\alpha$, by the
probability of returning to $\beta$ infinitely often, starting
from~$\beta$.

\vfill

\begin{lem}
  \label{lem:7}
  Let $\alpha,\beta$ be two global states, and let
  $(R^n_\beta)_{n\geq1}$ be the successive return times
  to~$\beta$. Let
  $h=\un{\bigcap_{n\geq1}\{R^n_\beta<\infty\}}$\,. Then:
  \begin{equation}
    \label{eq:40}
    \besp_\alpha(h)=\bpr_\alpha(R_\beta<\infty)\cdot\besp_\beta(h)\,. 
  \end{equation}
\end{lem}

\vfill

\proof
  Applying the Asynchronous Strong Markov property (Theorem~\ref{thr:2})
  with stopping time $R_\beta$ and function~$h$, we get:
  $\besp_\alpha\bigl(h\circ\theta_{R_\beta}|\FFF_{R_\beta}\bigr)=
  \besp_{\gamma(\omega_{R_\beta})}(h)$\,. The right side of this
  equality is simply the constant $\besp_\beta(h)$ on
  $\{R_\beta<\infty\}$. We multiply both sides by
  $\un{R_\beta<\infty}$, which is $\FFF_{R_\beta}$-measurable by
  definition of $\FFF_{R_\beta}$ and can therefore be put inside the
  $\besp_\alpha(\cdot|\FFF_{R_\beta})$ sign, to obtain:
\begin{equation*}
  \label{eq:39}
  \besp_\alpha(\un{R_\beta<\infty} h\circ\theta_{R_\beta}|\FFF_{R_\beta})=
\un{R_\beta<\infty}\besp_\beta(h).
\end{equation*}
We observe that $\un{R_\beta<\infty}h\circ\theta_{R_\beta}=h$, and
therefore
$\besp_\alpha(h|\FFF_{R_\beta})=\un{R_\beta<\infty}\besp_\beta(h)$. Taking
the $\besp_\alpha$-expectations yields identity~(\ref{eq:40}).
\qed

\vfill

\subsection{Recurrent and Transient Global States}
\label{sec:recurr-transc-glob}

In Markov chain theory, the Strong Markov property is a fundamental
tool for studying so-called recurrent states, those states to which
the chain returns infinitely often almost surely. There is a strong
parallel between 
%
Markov chain theory and this part of \abtcp\ theory:
recurrence concerns global states, and the infinite return is defined
through the successive return times defined
in~\S~\ref{sec:exampl-stopp-times}. And the Asynchronous Strong Markov
property is the fundamental tool in this study.

Denoting as in Definition~\ref{def:8} by $(R^n_\alpha)_{n\geq1}$ the
successive returns to $\alpha\in X_0$\,, we say that a global
trajectory $\omega\in\Omega$ \tdefine{returns infinitely often}
  to~$\alpha$ if $R^n_\alpha(\omega)<\infty$ for all integers
$n\geq1$.

\vfill

\begin{propdef}
  \label{prop:3}
  Let $(\bpr_\alpha)_{\alpha\in X_0}$ be a \abtcp. 
  \begin{enumerate}[\em(1)]
  \item\label{item:14} For any $\alpha\in
  X_0$, the set of trajectories that return infinitely often to
  $\alpha$ has\/ $\bpr_\alpha$-probability either $0$ or~$1$. Following
  Markov chain terminology, we will say that:
\begin{itemize}
\item $\alpha$ is \tdefine{recurrent} if\/
  $\bpr_\alpha\bigl(\bigcap_{n\geq1}\{R^n_\alpha<\infty\}\bigr)=1$,
  which is equivalent to: 
\begin{equation*}
\bpr_\alpha(R_\alpha<\infty)=1\,.
\end{equation*}
\item $\alpha$ is \tdefine{transient} if\/
  $\bpr_\alpha\bigl(\bigcap_{n\geq0}\{R^n_\alpha<\infty\}\bigr)=0$,
  which is equivalent to:
\begin{equation*}
\bpr_\alpha(R_\alpha<\infty)<1.
\end{equation*}
\end{itemize}
\item\label{item:15} There is at least one recurrent state in~$X_0$\,.
\item\label{item:16} If $\alpha$ is a recurrent state, then the
  successive returning trajectory to $\alpha$ defined by
  $\rho_n=\bigl(\theta_{R^{n-1}_\alpha}(\omega)\bigr)_{R_\alpha}$ for
  $n\geq1$, form a sequence of independent and identically distributed
  finite trajectories w.r.t.\ probability\/~$\bpr_\alpha$.
\item\label{item:17} If $\alpha$ is a recurrent state, and if $\beta$
  is reachable from~$\alpha$, then $\beta$ is recurrent and $\alpha$
  is reachable from~$\beta$.
  \end{enumerate}
\end{propdef}

\bigskip
\

\pagebreak
\proof
\
  \begin{enumerate}[(1)]
  \item The proof is adapted from~\cite[Proposition~1.2
    p.65]{revuz75}. Recall the usual transformation, for a measurable
    subset $A$ and some sub-\slgb\ $\GGG$ of a probability space
    $(\Omega,\FFF,\bpr)$:
    $\bpr(A)=\besp(\mathbf{1}_A)=\besp\bigl(\besp(\mathbf{1}_A|\GGG)\bigr)$.
    Putting $R=R_\alpha$ and $R^n=R^n_\alpha$, we apply this
    transformation to $(\Omega,\FFF,\bpr_\alpha)$ with
    $A=\{R^n<\infty\}$ and $\GGG=\FFF_{R^{n-1}}$:
    \begin{align*} 
\bpr_\alpha(R^n<\infty)
	&=\besp_\alpha\bigl(
\besp_\alpha(\un{R^n<\infty}|\FFF_{R^{n-1}})\bigr).     
    \end{align*}
    Since $R^n=R^{n-1}+R\circ\theta_{R^{n-1}}$ we have:
    $\un{R^n<\infty}=\un{R^{n-1}<\infty}\cdot
    \un{R\circ\theta_{R^{n-1}}<\infty}$\,. Since
    $\un{R^{n-1}<\infty}$ is
    $\FFF_{R^{n-1}}$-measurable, the usual property of
    conditional expectation yields:
    \begin{align}
\label{eq:30}
\bpr_\alpha(R^n<\infty)
	&=\besp_\alpha\bigl(
\un{R^{n-1}<\infty}\besp_{\alpha}(\un{R\circ\theta_{R^{n-1}}<\infty}|\FFF_{R^{n-1}})
\bigr).
    \end{align}
Applying the Asynchronous Strong Markov property (Theorem~\ref{thr:2})
with stopping time $R^{n-1}$ and function $\un{R<\infty}$ we have:
\begin{equation}
  \label{eq:41}
\besp_\alpha(\un{R\circ\theta_{R^{n-1}}<\infty}|\FFF_{R^{n-1}})=
\besp_{\gamma(\omega_{R^{n-1}})}(\un{R<\infty})\,. 
\end{equation}
Since $\gamma(\omega_{R^{n-1}})=\alpha$ on $\{R^{n-1}<\infty\}$,
multiplying both sides of~(\ref{eq:41}) by $\un{R^{n-1}<\infty}$ brings:
\begin{equation}
  \label{eq:42}
  \un{R^{n-1}<\infty}\besp_\alpha(\un{R\circ\theta_{R^{n-1}}<\infty}|\FFF_{R^{n-1}})=
\un{R^{n-1}<\infty}\besp_\alpha(\un{R<\infty})\,.
\end{equation}
We take the $\bpr_\alpha$-expectation of both sides of~(\ref{eq:42})
and report the result in~(\ref{eq:30}) to obtain:
\begin{equation}
  \label{eq:43}
  \bpr_\alpha(R^n<\infty)=\bpr_\alpha(R<\infty)\cdot\bpr_\alpha(R^{n-1}<\infty)\,.
\end{equation}
It follows from Borel-Cantelli Lemma that $\alpha$ is recurrent if
$\bpr_\alpha(R<\infty)=1$, and transient if $\bpr_\alpha(R<\infty)<1$.
\item Pick any $\alpha\in X_0$\,, and let $(R^n)_{n\geq1}$ be the
  successive return times to the square set~$X_0$
  (cf.~\S~\ref{sec:first-hitting-time-1}). With
  $\bpr_\alpha$-probability~$1$, $R^n<\infty$ for all $n\geq1$. It
  follows from Lemma~\ref{lem:6} applied with $A=X_0$ that, for some
  $\beta\in X_0$:
\begin{equation}
\label{eq:16}
\bpr_\alpha\Bigl(\bigcap_{n\geq1}\{ R^n_{\beta}<\infty\}\Bigr) >0.
\end{equation}

It remains to show that~(\ref{eq:16}) is still valid with
$\bpr_\beta$ in place of~$\bpr_\alpha$\,.  Let $h$ be the
non negative function
$h=\un{\bigcap_{n\geq1}\{R^n_\beta<\infty\}}$. Then 
 $\besp_\alpha(h)=\bpr_\alpha(R_\beta<\infty)\cdot \besp_\beta(h)$ by
 Lemma~\ref{lem:7}.  Since $\besp_\alpha(h)>0$ by
Eq.~(\ref{eq:16}), it follows that $\besp_\beta(h)>0$, showing that
$\beta$ is recurrent.

\item\label{itemiid} Observe that the $\rho_n$ are related to
  $R^n_\alpha$ through the identity:
  $\omega_{R^{n+1}_\alpha}=\omega_{R^n_\alpha}\cdot\rho_n$\,. Now, let
  $v_1,\dots,v_n$ be $n$ finite trajectories in the range
  of~$R_\alpha$\,. Since the $R^i_\alpha$ are stopping times, we have
  the equality $\{\rho_1=v_1,\dots,\rho_{n}=v_{n}\}=
  \{\omega_{R^{n}_\alpha}=v_1\cdot\ldots\cdot v_{n}\}=\uparrow
  (v_1\cdot\ldots\cdot v_{n})$\,. 
 The chain rule yields:
\begin{equation}
  \label{eq:25}
\bpr_\alpha(\rho_1=v_1,\dots,\rho_n=v_n)=
\bpr_\alpha(\uparrow v_1\cdot\ldots\cdot v_{n-1})\times\\
\bpr_\alpha\bigl(\uparrow v_1\cdot\ldots\cdot v_n|\uparrow
v_1\cdot\ldots\cdot v_{n-1}\bigr)\,.
\end{equation}

The Markov property~(\ref{eq:4}) combines with
$\gamma(v_1\cdot\ldots\cdot v_{n-1})=\alpha$ to rewrite the conditional
probability in Eq.~(\ref{eq:25}) as follows:
\[
\bpr_\alpha\bigl(\uparrow v_1\cdot\ldots\cdot v_n|\uparrow
v_1\cdot\ldots\cdot v_{n-1}\bigr)=\bpr_\alpha(\uparrow v_n)\,.
\]
Since $v_n$ is in the range of~$R_\alpha$\,, and since $R_\alpha$ is a
stopping time, $\uparrow v_n=\{R_\alpha=v_n\}$. We replace thus the
conditional probability in Eq.~(\ref{eq:25}) by
$\bpr_\alpha(R_\alpha=v_n)$, and apply $n$ times the same
transformation to finally obtain the identity:
\begin{equation*}
  \label{eq:18}
\bpr_\alpha(\rho_1=v_1,\dots,\rho_n=v_n)=\bpr_\alpha  (R_\alpha= v_1)
\cdot\ldots\cdot
\bpr_\alpha(R_\alpha= v_n)\,,
\end{equation*}
showing that the $\rho_n$ are i.i.d.\ random variables,  with the law
of~$R_\alpha$\,.

\item Consider the two measurable and non negative functions:
  \begin{equation*}
    h_\alpha=\un{\bigcap_{n\geq1}\{R^n_\alpha<\infty\}}\,,\qquad 
h_\beta=\un{\bigcap_{n\geq1}\{R^n_\beta<\infty\}}\,. 
  \end{equation*}
  Since the successive returns to $\alpha$ are i.i.d. by virtue of
  point~\ref{itemiid} above, each $\rho_n$ one has positive
  $\bpr_\alpha$-probability of hitting~$\beta$, otherwise the
  $\bpr_\alpha$-probability of ever hitting $\beta$ would be zero,
  contradicting the assumption that $\beta$ is reachable
  from~$\alpha$. Hence, by Borel-Cantelli Lemma,
  $\besp_\alpha(h_\beta)>0$.  Since
  $\besp_\alpha(h_\beta)=\bpr_\alpha(R_\beta<\infty)\cdot\besp_\beta(h_\beta)$
  by Lemma~\ref{lem:7}, this implies that $\besp_\beta(h_\beta)>0$ and
  thus $\beta$ is recurrent.

  To prove that $\alpha$ is reachable from~$\beta$, we apply the
  Asynchronous Strong Markov property (Theorem~\ref{thr:2}) with stopping
  time $\FFF_{R_\beta}$ and function~$h_\alpha$\,. We then multiply
  the resulting identity by $\un{R_\beta<\infty}$, and take into
  account that $\un{R_\beta<\infty}$ is $\FFF_{R_\beta}$-measurable on
  the one hand, and that $\gamma(\omega_{R_\beta})=\beta$ on
  $\{R_\beta<\infty\}$ on the other hand to obtain:
  \begin{equation}
\label{eq:21}
\besp_\alpha\bigl(
\un{R_\beta<\infty}h_\alpha\circ\theta_{R_\beta}|\FFF_{R_\beta}\bigr)
=\un{R_\beta<\infty}\cdot 
\besp_\beta(h_\alpha)\,.
\end{equation}
Observe that $h_\alpha\circ\theta_{R_\beta}=h_\alpha$ on
$\{R_\beta<\infty\}$, therefore the following identity is valid
everywhere:
\mbox{$\un{R_\beta<\infty}h_\alpha\circ\theta_{R_\beta}=\un{R_\beta<\infty}h_\alpha$}\,. By
assumption, $\alpha$~is recurrent, hence $h_\alpha=1$
$\bpr_\alpha$-almost surely, and finally
$\un{R_\beta<\infty}h_\alpha\circ\theta_{R_\beta}=\un{R_\beta<\infty}$
$\bpr_\alpha$-almost surely. Replacing thus
\mbox{$\un{R_\beta<\infty}h_\alpha\circ\theta_{R_\beta}$} by
$\un{R_\beta<\infty}$ in Eq.~(\ref{eq:21}), and taking the
$\bpr_\alpha$-expectations of both sides yields:
\[
\bpr_\alpha(R_\beta<\infty)=\bpr_\alpha(R_\beta<\infty)\cdot\besp_\beta(h_\alpha)\,.
\]
But $\beta$ is assumed to be reachable from~$\alpha$, hence
$\bpr_\alpha(R_\beta<\infty)>0$, and thus $\besp_\beta(h_\alpha)=1$,
implying in particular that $\alpha$ is reachable
from~$\beta$. \qed
\end{enumerate}

\subsection{Irreducible Components}
\label{sec:irred-comp}

With the notion of recurrent state at hand, it is now possible to
introduce the notions of irreducible process and irreducible components
of a \abtcp. 

\begin{defi}
  \label{def:11}
  Let\/ $\pr=(\bpr_\alpha)_{\alpha\in X_0}$ be a \abtcp. We say that\/
  $\pr$ is \tdefine{irreducible} if every $\alpha\in X_0$ is reachable
  from every $\beta\in X_0$\,.
\end{defi}

\begin{prop}
  \label{prop:5}
If a \abtcp\ is irreducible, then every global state is recurrent.
\end{prop}

\proof
 By Proposition~\ref{prop:3}, point~\ref{item:15}, there is some
 recurrent state $\alpha\in X_0$. But then, since any $\beta\in X_0$
 is reachable from~$\alpha$, $\beta$~is recurrent by
 point~\ref{item:17} of the same proposition. 
\qed

The result in Proposition~\ref{prop:9} below says that the study of
Markov \tcp es essentially reduces to the study of irreducible
processes, especially if one is interested in asymptotic properties
(so-called limit theorems from probability theory such as the Law of
Large Numbers or the Central Limit Theorem). For this we use the
notion of subprocess introduced in Definition~\ref{def:1}, and introduce
irreducible components for \abtcp s.

\pagebreak

\begin{defi}
  \label{def:16}
  An \tdefine{irreducible component} of a probabilistic \tcp\/
  $\pr=(\bpr_\alpha)_{\alpha\in X_0}$ is any subset $X_1\subset X_0$
  such that, for all $\alpha\in X_1$~:
  \begin{enumerate}[(1)]
  \item\label{item:3} any $\beta\in X_1$ is reachable from~$\alpha$;
    and
  \item\label{item:4} if $\beta\in X_0$ is reachable
    from~$\alpha$, then $\beta\in X_1$\,.
  \end{enumerate}
\end{defi}

Point~\ref{item:4} in Definition~\ref{def:16} ensures that
$(\bpr_\alpha)_{\alpha\in X_1}$ is indeed a probabilistic process
(Definition~\ref{def:1}). Therefore, if $X_1$ is an irreducible
component of \abtcp\ $\pr=(\bpr_\alpha)_{\alpha\in X_0}$, then the
family $(\bpr_\alpha)_{\alpha\in X_1}$ forms a subprocess of~$\pr$,
which is obviously an irreducible \abtcp. It follows from
Proposition~\ref{prop:5} that any element $\alpha$ of an irreducible
component is recurrent.  Any two irreducible components are disjoint.
Finally, if $\alpha$ is recurrent, then $\alpha$ belongs to a unique
irreducible component, namely the set $X_1$ of those $\beta$ which are
reachable from~$\alpha$ (the fact that $X_1$ is indeed an irreducible
component follows from Proposition~\ref{prop:3}). Since recurrent
states exist by Proposition~\ref{prop:3}, this implies that any
\abtcp\ has at least one irreducible component.

\begin{prop}
  \label{prop:9}
  If\/ $\pr=(\bpr_\alpha)_{\alpha\in X_0}$ is a \abtcp, there exists a
  stopping time $T:\Omega\to\TT$ such that $T$ is almost surely finite
  and $\gamma(\omega_T)$ belongs to some irreducible component
  of\/~$\pr$.
\end{prop}

\proof
  We fix an initial state $\alpha\in X_0$\,.  Let $(R^n)_{n\geq1}$
  denote the successive return times to the square set~$X_0$
  (cf.~\S~\ref{sec:first-hitting-time-1}). As already observed several
  times, $\bpr_\alpha(R^n<\infty)=1$ for all $n\geq1$, and therefore,
  if we put $B=\bigl\{\beta\in X_0\tq
  \bpr_\alpha\bigl(\bigcap_{n\geq1}R^n_\beta<\infty\}\bigr)>0\bigr\}$\,,
  it follows from Lemma~\ref{lem:6} that:
\begin{equation}
\label{eq:46}
\Omega=\bigcup_{\beta\in
  B}\bigcap_{n\geq1}\{R^n_\beta<\infty\}\,.
\end{equation}
Pick exactly one global state $\alpha_i$ for each irreducible
component. Let $T_i$ be the first hitting time of~$\alpha_i$, and put:
\begin{equation}
  \label{eq:36}
  \notag
  \forall\omega\in\Omega,\quad\omega_T=\inf_i\{\omega_{T_i}\}.
\end{equation}
For each $\beta\in B$, let $\betatilde$ be the unique recurrent state
$\alpha_i$ of the same irreducible component. Then $\betatilde$ is
reachable from~$\beta$, and therefore:
\begin{equation}
\label{eq:47}
  \omega\in\bigcap_{n\geq1}\{R^n_\beta<\infty\}\Rightarrow 
R_{\betatilde}(\omega)<\infty\quad\text{$\bpr_\alpha$-a.s.}
\end{equation}
>From Eqs.~(\ref{eq:46})(\ref{eq:47}) we deduce that $\omega_T<\infty$
$\bpr_\alpha$-almost surely. Hence, on the one hand, at least one
$T_i$ is finite $\bpr_\alpha$-almost surely.  On the other hand, only
one of them is finite, since the $\alpha_i$ have been chosen in
different irreducible components. Therefore $\omega_T=\omega_{T_i}$
for some~$T_i$, and thus $\gamma(\omega_T)$ does belong to some
irreducible component, as claimed.
\qed

\subsection{Open and Closed Markov Two-Components Processes}
\label{sec:open-closed-tcp}

Besides the classical application of the Strong Markov Property to
recurrence and transience, it also applies to the notion of open and
closed processes which is specific to the two-components
framework. Open and closed processes have been defined in
Definition~\ref{def:17}.

\begin{prop}
  \label{prop:6}
Let\/ $\pr=(\bpr_\alpha)_{\alpha\in X_0}$ be a \abtcp.
\begin{enumerate}[\em(1)]
\item\label{item:18} Let  $\alpha\in X_0$ be a recurrent state.  Then a
  global trajectory $\omega$ synchronizes infinitely often with\/
  $\bpr_\alpha$-probability\/ $1$ if at least some synchronization
  state is reachable from~$\alpha$,  and
  with\/ $\bpr_\alpha$-probability\/ $0$ otherwise.
\item If\/ $\pr$ is irreducible, then\/ $\pr$ is closed or open.
\end{enumerate}
\end{prop}

\proof\
  \begin{enumerate}[(1)]
  \item\label{item:9} Let $R_Q$ be first return time to the square set
    $X_0\cap(Q\times Q)$ (see~\S~\ref{sec:first-hitting-time-1}), and
    consider the stopping time $U=R_Q+R_\alpha\circ\theta_{R_Q}$\,,
    corresponding to reaching $\alpha$ after having reached~$Q$. This
    is indeed a stopping time by virtue of Lemma~\ref{lem:8}.  Let
    $(U_n)_{n\geq1}$ be the iterated stopping times associated with
    $U$ as in Definition~\ref{def:15}. If $h_n=\un{U_n<\infty}$, the same
    technique involving the Asynchronous Strong Markov property
    (Theorem~\ref{thr:2}) than in the proof of Proposition~\ref{prop:3},
    point~\ref{item:14}, shows that:
    $\besp_\alpha(h_n|\FFF_{U_{n-1}})=h_{n-1}\besp_\alpha(\un{U<\infty})$. Therefore,
    if $a=\bpr_\alpha(U<\infty)$ one has
    $\bpr_\alpha(U_n<\infty)=a^n$\,. Since $\alpha$ is recurrent,
    Proposition~\ref{prop:3}, point~\ref{item:14} implies that $R_\alpha$ is
    $\bpr_\alpha$-almost surely finite, hence
    \mbox{$\bpr_\alpha(U<\infty)=\bpr_\alpha(R_Q<\infty)$}. Since a
    trajectory $\omega$ synchronizes infinitely often if and only if
    $U_n<\infty$ for all $n\geq1$, Borel-Cantelli Lemma implies that
    $\omega$ has $\bpr_\alpha$-probability $1$ of synchronizing
    infinitely often if $\bpr_\alpha(R_Q<\infty)=1$, and $0$
    otherwise.

    It remains to show that $\bpr_\alpha(R_Q<\infty)=1$ if and only if
    some state of the form $(x,x)$ with $x\in Q$ is reachable
    from~$\alpha$. Since $R_Q=\bigwedge_{x\in Q}R_{(x,x)}$\,,
    obviously if no $(x,x)$ is reachable from $\alpha$ then
    $\bpr_\alpha(R_Q<\infty)=0$. Conversely, assume that some $(x,x)$
    with $x\in Q$ is reachable from~$\alpha$. Then $(x,x)$ is
    recurrent, by point~\ref{item:17} of Proposition~\ref{prop:3}, and
    Lemma~\ref{lem:7} implies that $R_{(x,x)}<\infty$
    $\bpr_\alpha$-almost surely. But $R_Q\leq R_{(x,x)}$, hence
    $\bpr_\alpha(R_Q<\infty)=1$, as claimed.

  \item If $\pr$ is irreducible, then by Proposition~\ref{prop:3},
    every $\alpha\in X_0$ is recurrent, therefore point~\ref{item:9}
    above applies to any $\alpha\in X_0$. Assume that the
    $\bpr_\alpha$-probability of synchronizing infinitely often is~$0$
    for some $\alpha\in X_0$\,, and let $\beta\in X_0$\,. Consider a
    finite trajectory $v$ such that $\bpr_\alpha(v)>0$ and
    $\gamma(v)=\beta$; such a $v$ exists since any $\beta$ is
    reachable from~$\alpha$. Then $\bpr_\alpha$-\as\ every trajectory
    $\omega\in\uparrow v$ has no synchronization. But the
    $\bpr_\alpha$ probability measure on $\uparrow v$ coincides, up to
    the factor $\bpr_\alpha(v)\neq0$, with $\bpr_\beta$
    on~$\Omega$. Hence $\bpr_\beta$-\as\ every $\omega\in\Omega$ has
    no synchronization, and since this is true for every $\beta\in
    X_0$, the process $\pr$ is open. The same method applies to show
    that $\pr$ is closed if the probability of synchronizing
    infinitely often is $1$ for some $\alpha\in X_0$. This concludes
    the proof.\qed
  \end{enumerate}

\section{The Local Independence Property}
\label{sec:prop-mark-two}

Having adapted Markovian concepts from Markov chain theory, we now
focus on a topic specific to the asynchronous framework, without
equivalent in Markov chain theory: the probabilistic correlation
between private behaviors of local components. It is desirable to
have a kind of probabilistic independence between private parts of
trajectories: otherwise, hidden synchronization constraints would be
encoded in the probabilistic structure, while we expect
synchronization to occur only on explicit synchronization
states. Probabilistic independence of random variables $\omega^1$ and
$\omega^2$ however is too much to ask; their synchronization is an
obstacle to their mere probabilistic independence. This is easy to
understand from an information theoretic viewpoint: the knowledge of
$\omega^1$ gives indeed information on~$\omega^2$, since it precisely
determines the $Q$-sequence of~$\omega^2$.  The weaker notion of
conditional independence proves to be adapted to our purpose. The
Local Independence Property that we introduce informally states that
the two local components have the maximal probabilistic independence
they can have, considering their natural synchronization constraints.

\medskip

Recall that $Y=\seq Yn$ has been defined in Definition~\ref{def:17} as the
$Q$-sequence induced by some trajectory $\omega\in\Omega$, to which we
have added $Y_{-1}=*$ and $Y_n=*$ for large $n$ if the $Q$-sequence is
finite, for some fixed specified value~$*$.  We proceed in a
similar way to define the sequence $\seq\sigma n$ of random elementary
trajectories, referring to the decomposition of a trajectory $\omega$
as a concatenation of elementary trajectories from
Proposition~\ref{prop:1}. If $\sigma_n$ is defined only until some
integer~$N$ (that is, in case \ref{item:2} of Proposition~\ref{prop:1}) we
define $\sigma_{N+1}$ as the synchronization free trajectory such that
$\omega=\sigma_1\cdot\ldots\sigma_N\cdot\sigma_{N+1}$ and $\sigma_n=*$
for $n>N+1$.



Then we observe the following property:

\begin{prop}
  \label{prop:7}
  Let\/ $\pr$ be the synchronization product of two Markov
  chains. Decomposing $\sigma_n$ as
  $\sigma_n=(\sigma^1_n,\sigma^2_n)$ we have: for all $\alpha\in X_0$
  and for every integer $n\geq0$, $\sigma^1_n$~and $\sigma^2_n$ are
  two random variables independent conditionally on the pair
  $(Y_{n-1},Y_n)$ with respect to\/~$\bpr_\alpha$\,.
\end{prop}

\proof
  Since $\pr$ satisfies the Markov property, the statement is equivalent
  to the independence of $\sigma_n^1$ and~$\sigma^2_n$, conditionally
  on~$Y_n$, and with respect to~$\bpr_{Y_{n-1}}$\,. But this follows
  from the construction of the law of
  $\sigma_n=(\sigma^1_n,\sigma^2_n)$ given
  in~\S~\ref{sec:synchr-two-mark}.
\qed

In order to generalize the above property to processes which may
not be closed, and at the cost of a little more abstraction, we
introduce the following definition.

\begin{defi}
  \label{def:12}
  Let\/ $\pr=(\bpr_\alpha)_{\alpha\in X_0}$ be a \abtcp, let\/ $Y$ be
  the associated random synchronization sequence. Let $\omega^1$ and
  $\omega^2$ denote the local components of global trajectories, so
  that $\omega=(\omega^1,\omega^2)$ for $\omega\in\Omega$. We say
  that\/ $\pr$ has the \tdefine{local independence property}
  (abbreviated {\normalfont\tdefine{LIP}}) if\/ $\omega^1$ and\/
  $\omega^2$ are independent conditionally\footnote{Recall that two
    random variables $X_1$ and $X_2$ are independent w.r.t.\ a
    \slgb~$\GGG$ if
    $\besp(\varphi_1\cdot\varphi_2|\GGG)=\besp(\varphi_1|\GGG)
    \cdot\besp(\varphi_2|\GGG)$, for all non negative and bounded
    variables $\varphi_1$ and~$\varphi_2$, measurable with respect to
    $X_1$ and to $X_2$
    respectively. See~\textsl{e.g.}~\cite[Chapter~IV]{neveu65}. Here,
    the independence conditionally to $Y$ means the independence
    w.r.t.\ the \slgb\ $\langle Y\rangle$ generated by~$Y$.}  to~$Y$
  with respect to\/~$\bpr_\alpha$\,, for all \mbox{$\alpha\in X_0$}\,.
\end{defi}

The following theorem relates this definition with the previous
property stated in Proposition~\ref{prop:7} for the synchronization of
Markov chains. 

\begin{thm}
  \label{thr:3}
  Let\/ $\pr=(\bpr_\alpha)_{\alpha\in X_0}$ be a \abtcp. Then\/ $\pr$
  satisfies the {\normalfont LIP} if and only if the random variables
  $\sigma_n^1$ and $\sigma_n^2$ are independent conditionally on the
  pair $(Y_{n-1},Y_n)$, with respect to\/ $\bpr_\alpha$ for all
  $n\geq0$ and for all $\alpha\in X_0$\,.
\end{thm}

\proof
  Let $(a)$ be the property that $\omega^1$ and $\omega^2$ are
  independent conditionally on~$Y$, and let $(b)$ be the property
  stated in the theorem.

  \medskip \textsl{Proof of $(a)\Rightarrow(b)$.}\quad Thanks to the
  Markov property, it is enough to consider \mbox{$n=1$}. We denote 
  $\sigma^1_1$ and $\sigma^2_1$ by $\sigma^1$ and~$\sigma^2$\,, and we
  put: $Z^1=\besp_\alpha\bigl(\un{\sigma^1=z^1}\,|\,Y\bigr)$\,,
  $Z^2=\besp_\alpha\bigl(\un{\sigma^2=z^2}\,|\,Y\bigr)$\,, and $
  Z=\besp_\alpha\bigl(\un{\sigma^1=z^1,\,\sigma^2=z^2}\,|\,Y\bigr)$\,.
  These three random variables are constant on $\{Y_1=b\}$ and,
  by~$(a)$, satisfy $Z=Z^1\cdot Z^2$, whence:
\begin{align*}
  \bpr_\alpha(\sigma^1=z^1,\,\sigma^2=z^2|Y_1=b)&= Z\big|_{\{Y_1=b\}}\\
&=  Z^1\big|_{\{Y_1=b\}}\cdot  Z^2\big|_{\{Y_1=b\}}\\
&=\bpr_\alpha(\sigma^1=z^1\,|Y_1=b)\times \bpr_\alpha(\sigma^2=z^2|Y_1=b)\,,
\end{align*}
as expected.

\medskip \textsl{Proof of $(b)\Rightarrow(a)$.}\quad From $(b)$ used
in conjunction with the Markov property and the chain rule, we get for
integers $m\geq n$ and with short notations:
\begin{multline}
  \label{eq:33}
  \notag
\bpr_\alpha\bigl(\sigma^1_1,\ldots,\sigma^1_m,\sigma^2_1,\ldots,\sigma^2_m\big|
Y_1,\ldots,Y_n\bigr)=
\bpr_\alpha\bigl(\sigma^1_1,\ldots,\sigma^1_m\big|
Y_1,\ldots,Y_n\bigr)\times\\
\bpr_\alpha\bigl(\sigma^2_1,\ldots,\sigma^2_m\big|
Y_1,\ldots,Y_n\bigr).
\end{multline}
The \slgb\ generated by the random trajectories
$(\sigma^i_k,\,k\geq 1)$ for $i=1,2$ coincides with the \slgb\
generated by~$\omega^i$, since $\omega^i$ is obtained as the
concatenation of these---the concatenation being finite or infinite.
Hence, for any bounded non negative and measurable functions $h^1$
and~$h^2$:
\begin{equation*}
  \besp_\alpha\bigl(
  h^1(\omega^1)\cdot h^2(\omega^2)\big|
Y_1,\ldots,Y_n\bigr)=
\besp_\alpha\bigl(h^1(\omega^1)\big|
Y_1,\ldots,Y_n\bigr)
\cdot
\besp_\alpha\bigl(h^2(\omega^2)\big| Y_1,\ldots,Y_n\bigr).
\end{equation*}
The sequence of \slgb s $\langle Y_1,\ldots,Y_n\rangle$ is increasing,
and converges to~$\langle Y\rangle$. Therefore by the special case
\cite[Theorem~35.6 p.470]{billingsley95} of the Martingale convergence
theorem, we get by taking the limit $n\to\infty$:
\begin{equation}
  \label{eq:35}
  \notag
  \besp_\alpha\bigl(
  h^1(\omega^1)\cdot h^2(\omega^2)\big|Y\bigr)=
\besp_\alpha\bigl(h^1(\omega^1)\big|Y\bigr)
\cdot
\besp_\alpha\bigl(h^2(\omega^2)\big|Y\bigr),
\end{equation}
completing the proof.
\qed

\begin{cor}
  \label{cor:1}
The synchronization product of Markov chains satisfies the
{\normalfont LIP}.
\end{cor}

Having the specified value $*$ assigned to some $Y_{k}$ and $\sigma_k$
described above has the following effect with regard to
Theorem~\ref{thr:3}: the statement is trivial if both $\sigma_k$,
$Y_{k-1}$ and $Y_k$ assume their constant values~$*$; but it implies
the probabilistic independence of $\sigma^1_{N+1}$ and
$\sigma^2_{N+1}$ with respect to~$\bpr_{Y_{N}}$\,, where $N$ is the
last synchronization index. In other words, the local trajectories are
independent after their last synchronization.

It is useful to examine a degenerated case of Definition~\ref{def:12},
where the conditional independence reduces to probabilistic
independence. 

\begin{lem}
  \label{lem:2}
  Let\/ $\pr=(\bpr_\alpha)_{\alpha\in X_0}$ be a \abtcp. Let\/
  $\omega^1$ and\/ $\omega^2$ denote the local components of global
  trajectories. Assume that, with respect to\/ $\bpr_\alpha$ for some
  state $\alpha\in X_0\,$, the two components\/ $\omega^1$ and\/
  $\omega^2$ are independent. Then $\omega^1$ and\/ $\omega^2$ are the
  sample paths of two independent Markov chains.
\end{lem}

\proof
  Fix $\alpha\in X_0$, and for each $i=1,2$ let $P^i$ denote the law
  of~$\omega^i$, characterized by $P^i(\omega^i\geq
  s^i)=\bpr_\alpha(\omega^i\geq s^i)$, with $s^i$ ranging over the
  finite local trajectories on site~$i$. We show that the conditional
  law $P^i(s^i\cdot\bullet|\uparrow s^i)$ only depends on the last
  state of~$s^i$, which is enough to obtain that $\omega^i$ follows
  the law of a homogeneous Markov chain. Consider $i=1$, the case
  $i=2$ is identical. Consider $s^1$ a finite sequence in~$S^1$ such
  that $P^1(\omega^1\geq s^1)>0$. It implies that there exists some
  sequence in~$S^2$, say~$s^2$, such that
  $\bpr_\alpha\bigl(\uparrow(s^1,s^2)\bigr)>0$. Put $s=(s^1,s^2)$ and
  let $(x^1,x^2)=\gamma(s)$. For any finite sequence $\sigma$
  in~$S^1$, we have:
  \begin{align}
\label{eq:29}
\notag
    P^1(\omega^1\geq s^1\cdot\sigma|\omega^1\geq s^1)&=
\frac{\bpr_\alpha(\omega^1\geq
  s^1\cdot\sigma)}{\bpr_\alpha(\omega^1\geq s^1)}\\
\notag
&=\frac{\bpr_\alpha(\omega^1\geq
  s^1\cdot\sigma,\,\omega^2\geq s^2)}{\bpr_\alpha(\omega^1\geq
  s^1,\,\omega^2\geq s^2)}&\text{by independence}\\
&=\bpr_{(x^1,x^2)}\bigl(\omega^1\geq\sigma \bigr).
  \end{align}
Obviously, the expression $P^1(\omega^1\geq
s^1\cdot\sigma|\omega^1\geq s^1)$ does not depend on~$x^2$, since
$x^2$ is the last state of the arbitrary chosen
sequence~$s^2$. Therefore, the right member of~\eqref{eq:29} does not
depend on $x^2$ neither, hence it only depends on $x^1$ and~$\sigma$,
which was to be proved.
\qed

\begin{prop}
  \label{prop:8}
  Let\/ $\pr=(\bpr_\alpha)_{\alpha\in X_0}$ be a \abtcp\ with the
  {\normalfont LIP}. Let\/ $\omega^1$ and\/ $\omega^2$ denote the
  local components of global trajectories.
  \begin{enumerate}[\em(1)]
  \item If $Q=\emptyset$, then $\omega^1$ and $\omega^2$ are two
    independent Markov chains, with respect to\/~$\bpr_\alpha$ for any
    $\alpha\in X_0$.
  \item\label{item:5} If $Q$ is a singleton, and if $\alpha$ is a
    recurrent state, then $\omega^1$ and $\omega^2$ are two
    independent Markov chains, with respect to\/~$\bpr_\alpha$.
  \end{enumerate}
\end{prop}

\proof\
  \begin{enumerate}[(1)]
  \item\label{item:19} Since $Q=\emptyset$, the synchronization
    sequence $Y$ is constant, $Y=(*,*,*,\ldots)$. The conditional
    independence in the definition of the LIP reduces to probabilistic
    independence. The result follows then by Lemma~\ref{lem:2}.
  \item Let $Q=\{\beta\}$. By Proposition~\ref{prop:6},
    point~\ref{item:18}, $\omega$~synchronizes infinitely often with
    $\bpr_\alpha$-probability either $0$ or~$1$. If it is with
    probability~$0$, then $Y=(*,*,*,\ldots)$ $\bpr_\alpha$-\as, and the
    same method than in point~\ref{item:19} above applies. If it is
    with probability~$1$, then $Y$ is still constant, now
    $Y=(*,\beta,\beta,\beta,\ldots)$. The same method applies again.\qed
  \end{enumerate}

\begin{cor}
  \label{cor:2}
  An open \abtcp\ with the {\normalfont LIP} identifies with two
  independent homogeneous Markov chains.
\end{cor}

\section{Characterization of Markov Two-Components 
Processes with the LIP}
\label{sec:gener-constr-mark}

The topic of this section is to characterize a \abtcp\ with the LIP by
means of a finite family of real numbers, very much as the transition
matrix of a Markov chain does. It turns out that the law of a \abtcp\
with the LIP is entirely specified by a finite family of transition
matrices.  We will also investigate, conversely, if such a family of
transition matrices always induces a \abtcp\ with the LIP, providing a
more general way of constructing \abtcp s than the synchronization
product of Markov chains. We show through a numerical example at the
end of the section that not any \abtcp\ can be obtained as the
synchronization product of two Markov chains.

\subsection{Technical Preliminaries}

We begin with two lemmas.

\begin{lem}
  \label{lem:3}
  Let\/ $\pr=(\bpr_\alpha)_{\alpha\in X_0}$ be a closed \abtcp, and let $Y$
  denote the associated synchronization sequence. Then for any
  $\alpha\in X_0$, $Y$~is a homogeneous Markov chain with respect
  to\/~$\bpr_\alpha$.
\end{lem}

\proof
  The formulation of Definition~\ref{def:3} applies to $Y$ as follows: for
  any two finite sequences $s$ and $u$ in~$Q$, the conditional
  probability $\bpr_\alpha(Y\geq s\cdot u|Y\geq s)$ only depends on
  $u$ and on the last state of~$s$. This shows that $Y$ is a
  homogeneous Markov chain. 
\qed

\begin{lem}
  \label{lem:4}
  Let\/ $\pr=(\bpr_\alpha)_{\alpha\in X_0}$ be a \abtcp\ with the
  {\normalfont LIP}, let $Y$ denote the associated synchronization
  sequence, and let $\seq \sigma n$ denote the sequence of elementary
  trajectories that decompose global trajectories
  (see~\S~{\normalfont\ref{sec:prop-mark-two}}).

  Then for every $n\geq0$ and for $i=1,2$, the sequence of states that
  appear in $\sigma^i_n$ is a stopped Markov chain with respect to the
  conditional probability $\bpr_\alpha(\,\cdot\,|Y_{n-1},Y_{n})$.
\end{lem}

\proof
  By the Markov property, there is no loss of generality in assuming
  that $n=0$. Using the notation $\sigma^i=\sigma^i_0$ for short, we
  thus have to prove that $\sigma^i$ is a stopped Markov chain with
  respect to $\bpr_{\alpha}(\,\cdot\,|Y_{0}=y)$, for any value $y\in
  Q$. We consider $i=1$ only, the case $i=2$ is similar. Let
  $(X_1,\dots,X_\tau)$ be the sequence of states in~$\sigma^1$, and
  let $\bprq$ denote the conditional probability
  $\bprq=\bpr_\alpha(\,\cdot\,|Y_0=y)$. Let $x_1,\dots,x_n$ be values
  in $S^1\setminus Q$, let $x_{n+1}\in Q\cup\{y\}$, and put
  $\delta=\bprq(X_{n+1}=x_{n+1}|X_1=x_1,\dots,X_n=x_n)$. We claim that
  $\delta$ only depends on $x_n$ and~$x_{n+1}$. Put $\alpha=(x,z)$. We
  calculate:
  \begin{align*}
    \delta&=\frac{\bprq(X_1=x_1,\dots, X_{n+1}=x_{n+1})}
{\bprq(X_1=x_1,\dots, X_{n}=x_{n})}\\
&=\frac{\bpr_\alpha(X_1=x_1,\dots, X_{n+1}=x_{n+1},X_\tau=y)}
{\bpr_\alpha(X_1=x_1,\dots, X_{n}=x_{n},X_\tau=y)}\,.
\end{align*}
We can rephrase $\{X_1=x_1,\dots, X_{n+1}=x_{n+1}\}$ in the
two-components framework as \mbox{$\{\omega^1\geq (x_1\cdot\ldots\cdot
  x_{n+1})\}=\uparrow (x_1\cdot\ldots\cdot x_{n+1},\epsilon)$},
observing that $(x_1\cdot\ldots\cdot x_{n+1},\epsilon)$ is indeed a
trajectory. The same applies to
$\{X_1=x_1,\ldots,X_n=x_n\}=\uparrow(x_1\cdot\ldots\cdot
x_n,\epsilon)$. Therefore the calculation continues as follows:
\begin{align*}
\delta&=\frac{\bpr_\alpha\bigl(\omega^1\geq(x_1\cdot\ldots\cdot x_{n+1}),\,X_\tau=y\bigr)}{\bpr_\alpha\bigl(\uparrow(x_1\cdot\ldots\cdot x_{n},\epsilon),\,X_\tau=y\bigr)}\\
&=\frac{\bpr_\alpha\bigl(\omega^1\geq(x_1\cdot\ldots\cdot
  x_{n+1}),\;X_\tau=y
\big|\uparrow(x_1\cdot\ldots\cdot x_n,\epsilon)\bigr)}
{\bpr_\alpha\bigl(\uparrow(x_1\cdot\ldots\cdot x_n,\epsilon),\, X_\tau=y
\big|\uparrow(x_1,\cdot\ldots\cdot x_n,\epsilon)\bigr)}\\
&=\frac{\bpr_{(x_n,z)}(\omega^1\geq x_{n+1},\,X_\tau=y)}
{\bpr_{(x_n,z)}(X_\tau=y)}=\bpr_{(x_n,z)}(\omega^1\geq x_{n+1}\big|X_\tau=y).
  \end{align*}
  On the last expression, it is clear that $\delta$ only depends on
  $x_n$ and~$x_{n+1}$, and not on $x_1,\dots,x_n$, showing our
  claim. This is enough to imply that $X_1,\dots,X_\tau$ are the terms
  of a homogeneous Markov chain.
\qed

\subsection{Adapted Family of Transition Matrices}
\label{sec:adpat-family-trans}

The two above lemmas suggest the following construction for \abtcp\
with the LIP. First consider a Markov chain $Y$ on the set of shared
states; then for any two consecutive values $y_{n-1}$ and $y_n$
of~$Y$, consider two independent stopped Markov chains $\sigma^1_n$
and~$\sigma^2_n$, with $\sigma^i_n$ taking values in
$\{y_n\}\cup(S^i\setminus Q)$, that reaches $y_n$ with probability one
and which is stopped at the first hitting time of~$y_n$. This
description is formalized in Theorem~\ref{thr:4} below. It is first
convenient to introduce the following definition.

\begin{defi}
  \label{def:13}
  An \tdefine{adapted family of transition matrices} is given by two
  families $(R^i_y)_{y\in Q_0}$\,, one for each $i=1,2$ and with $Q_0$
  some subset of~$Q$, such that:
\begin{enumerate}[(1)]
\item For each $y\in Q_0$ and $i=1,2$, $R^i_y$ is a stochastic matrix
  on $\{y\}\cup(S^i\setminus Q)$;
\item With respect to the transition matrix~$R^i_y$\,, the state $y$ is
  reachable from any state in $S^i\setminus Q$.
\end{enumerate}
\end{defi}

\noindent Using this definition, the existence and uniqueness result concerning
\abtcp\ with the LIP states as follows. We focus on closed processes
only, as suggested by Proposition~\ref{prop:9}, Proposition~\ref{prop:6} and
Corollary~\ref{cor:2}.
\newpage

\begin{thm}
  \label{thr:4}
  Any closed \abtcp\/\ $\pr$ with the {\normalfont LIP} induces the
  following elements, that entirely characterize\/~$\pr$:
\begin{enumerate}[\em(1)]
\item A transition matrix $R$ on the set $Q$ of shared states, defined
  as the transition matrix of the synchronization sequence $Y$ from
  Definition~{\normalfont\ref{def:17}};
\item An adapted family of transition matrices $(R^i_y)_{y\in Q_0}$,
  for $i=1,2$, where $Q_0$ is the essential set of values of\/~$Y$.
  For each $i=1,2$, and for $y\in Q_0$\,, $R^i_y$~is the transition
  matrix of the Markov chain~$\sigma^i_n$ with respect to the
  conditional probability $\bpr_\alpha(\,\cdot\,|Y_{n-1},Y_n=y)$,
  which is independent of the integer $n$ and of $\alpha\in X_0$,
  provided it is defined for these values.
\end{enumerate}
\noindent
Conversely, given a set of global states 
\begin{equation}
\label{eq:19}
\notag
X_0\subset \bigl\{(x,z)\in S^1\times S^2\tq (x\in Q)\wedge(z\in
Q)\Rightarrow x=z\bigr\},
\end{equation}
such that the set
\begin{equation}
  \label{eq:37}
  \notag
Q_0=\{y\in Q\tq(y,y)\in X_0\}.
\end{equation}
is nonempty; and considering:
\begin{enumerate}[\em(1)]
\item a transition matrix $R$ on the set~$Q_0$; and
\item an adapted family of transition matrices $(R_y^i)_{y\in Q_0}$\,,
\end{enumerate}
then there exists a unique \abtcp\ with the {\normalfont LIP}, defined
on $X_0$ and inducing $R$ and~$(R_y^i)_{y\in Q_0}$\,. This \abtcp\ is
closed.
\end{thm}

\proof
  The first part of the theorem follows from Lemmas~\ref{lem:3}
  and~\ref{lem:4}. For the second part, assume that the considered
  data are given. The construction of the process $\pr$ is essentially
  the same as the construction of the synchronization product of
  Markov chains, therefore we omit the routine arguments showing the
  existence and uniqueness of~$\pr$. What we need to show is that the
  \tcp\ obtained is indeed a \abtcp\ with the LIP. The LIP is obvious
  from the construction of $\pr$ combined with Theorem~\ref{thr:3}, hence
  we focus on the Markov property. Since the process is closed by
  construction, we rely on Lemma~\ref{lem:1} for this. Hence, let
  $\alpha\in X_0$, let $t$ be any elementary trajectory and let $s$ be
  any finite trajectory. The proof then follows the same steps than
  the proof of Theorem~\ref{thr:1}:
  \begin{enumerate}[(1)]
  \item \textsl{Step~$1$: $s$ synchronization free.}\quad 
Then $s\cdot t$ is an elementary trajectory. Put
  $\alpha=(x_0,z_0)$, $\gamma(s)=(x_1,z_1)$ and $\gamma(t)=(y,y)$. We
  have: $\uparrow(s\cdot t)=\{\sigma_1^1=s^1\cdot
  t^1,\,\sigma^2_1=s^2\cdot t^2\}$. Let $\bprq^i_b$ denote the
  probability associated with the Markov chain starting from $b$ and
  with transition matrix~$R^i_y$, for $i=1,2$ and $b\in Q$. We compute
  using the independence conditionally on~$Y_1$:
  \begin{align*}
    \bigl(\bpr_\alpha\bigr)_s\bigl(\uparrow t\bigr)&=
    \bigl(\bpr_\alpha\bigr)_s\bigl(\uparrow t\wedge Y_1=y\bigr)\\
    &= \frac{\bpr_\alpha\bigl(\uparrow(s\cdot t)\big|Y_1=y
      \bigr)}{\bpr_\alpha\bigl(\uparrow s\big|Y_1=y
      \bigr)}\\
   &=\frac{\bprq^1_{x_0}\bigl(\uparrow(s^1\cdot t^1)\bigr)}
    {\bprq^1_{x_0}\bigl(\uparrow s^1\bigr)} \cdot
    \frac{\bprq^2_{z_0}\bigl(\uparrow(s^2\cdot t^2)\bigr)}
    {\bprq^2_{z_0}\bigl(\uparrow s^2\bigr)}\\
&=\bprq^1_{x_1}(\uparrow t^1)\cdot
\bprq^2_{z_1}(\uparrow t^2).
\end{align*}
The last quantity only depends on $(x_1,z_1)=\gamma(s)$ and~$t$. In
particular, as expected, we have $\bigl(\bpr_\alpha\bigr)_s\bigl(\uparrow
t\bigr)=\bpr_{\gamma(s)}(\uparrow t)$.
\item \textsl{Step~$2$: $s$ is any finite trajectory.}\quad Using
  Step~$1$, as in the proof of Theorem~\ref{thr:1}.\qed
\end{enumerate}

\subsection{A Numerical Example}
\label{sec:few-examples}

In this subsection, we show on an example how the synchronization
product of Markov chains is to be interprated in terms of an adapted
family of transition matrices. We show that not any \abtcp\ can be
obtained from the synchronization of two Markov chains.

\medskip
Let $S^1=\{a,b,\c,\d\}$ and $S^2=\{\c,\d,e,f\}$, and let two transition
matrices $M^1$ and $M^2$ on $S^1$ and $S^2$ respectively. Take for
instance:
\begin{align*}
M^1&=\begin{matrix}
  a\\[\smalldim]b\\[\smalldim]\c\\[\smalldim]\d
\end{matrix}
\begin{pmatrix}
\frac13&\frac13&\frac13&0\\[\smalldim]
\frac12&\frac18&\frac18&\frac14\\[\smalldim]
\frac12&0&\frac14&\frac14\\[\smalldim]
0&\frac12&\frac14&\frac14
\end{pmatrix}
&
M^2&=M^1=\begin{matrix}
  e\\[\smalldim]f\\[\smalldim]\c\\[\smalldim]\d
\end{matrix}
\begin{pmatrix}
\frac13&\frac13&\frac13&0\\[\smalldim]
\frac12&\frac18&\frac18&\frac14\\[\smalldim]
\frac12&0&\frac14&\frac14\\[\smalldim]
0&\frac12&\frac14&\frac14
\end{pmatrix}\,.
\end{align*}
The matrices contain $0$ in some places, but that will not harm.

\subsubsection*{Computation of the adapted family of transition matrices.}
\label{sec:comp-adapt-family}

We need to compute the matrices $R^1_{\c}=R^2_{\c}$ and
$R^1_{\d}=R^2_{\d}$\,. Matrix $R^1_{\c}$ is a stochastic matrix on
$\{a,b,\c\}$, and drives the subsystem on site~$1$, conditionally on
``next synchronization is~$\c$''. Referring to the construction
detailed in~\S~\ref{sec:synchr-two-mark}, $R^1_{\c}$~is simply
obtained as follows: starting from matrix~$M^1$, suppress all lines
and columns attached to states in $Q$ different from~$\c$, here, this
is only state~$\d$. Finally, renormalize each line to obtain a
stochastic matrix. The same process is applied to obtain~$R^1_{\d}$\,:
\begin{align*}
R^1_{\c}&=
\begin{matrix}
a\\[\smalldim]b\\[\smalldim]\c
\end{matrix}
\begin{pmatrix}
  \frac13&\frac13&\frac13\\[\smalldim]
\frac23&\frac16&\frac16\\[\smalldim]
\frac23&0&\frac13
\end{pmatrix}&
R^1_{\d}&=
\begin{matrix}
  a\\[\smalldim]b\\[\smalldim]\d
\end{matrix}
\begin{pmatrix}
  \frac12&\frac12&0\\[\smalldim]
\frac47&\frac17&\frac27\\[\smalldim]
0&\frac23&\frac13
\end{pmatrix}
\end{align*}
This construction implies that the lines obtained from matrices
$R^1_{\c}$ and $R^1_{\d}$ by deleting the lines and columns relative
to shared states are \emph{proportional}: $
\begin{pmatrix}
  \frac13&\frac13
\end{pmatrix}$ is proportional to $
\begin{pmatrix}
  \frac12&\frac12
\end{pmatrix}$, and $
\begin{pmatrix}
  \frac23&\frac16
\end{pmatrix}$ is proportional to $
\begin{pmatrix}
  \frac47&\frac17
\end{pmatrix}$. Indeed, the lines of $R_{\c}^1$ and $R_{\d}^1$ are
obtained by renormalization after extraction from the same transition
matrix~$M^1$\,. We deduce from this observation a way to construct a
\abtcp\ with the LIP not obtained as a synchronization product of
Markov chains. Replace for example the $b$ line of $R^1_{\c}$ by $
\begin{pmatrix}
  0&0&1
\end{pmatrix}$ and leave $R^1_{\d}$ unchanged. This corresponds to
some closed \abtcp\ with LIP according to Theorem~\ref{thr:4}, which
cannot be a synchronization product of Markov chains.

We have obtained: \emph{not every \abtcp\ with the {\normalfont LIP}
  can be obtained as the synchronization product of two Markov chains}.

\subsubsection*{Computation of the matrix of the synchronization chain.}
\label{sec:comp-matr-synchr}

It remains to compute the transition matrix of the chain
$Y=(Y_n)_{n\geq1}$\,, which involves the law of $X^1_{\tau^1}$
and~$X^2_{\tau^2}$\,, where $\tau^i$ are the first hitting times to
$Q$ of chains $X^1$ and $X^2$ respectively, which we do here ``by
hand''. For a general theory, see for instance~\cite[Ch.~XII
\S\S58--59 \emph{Entrance and exit laws},
p.262\textit{ff}]{dellacherie88}.

Denoting by $M^1_{x}$ the law of chain $X^1$ starting from~$x$, one
has: $M^1_{x}(X^1_{\tau^1}=\c)=\sum_w M^1_{x}(w)$\,, where $w$ ranges
over words of the form $w=v\cdot \c$, and $v$ is any word on
$\{a,b\}$. Therefore, if $q_k(x)$ denotes, for any integer $k\geq0$:
\[
q_k(x)=\sum_{l_1,\ldots,l_k\in\{a,b\}}M^1_{x}(l_1\cdot\ldots\cdot
l_k\cdot \c)\,,
\]
one has $M^1_{x}(X^1_{\tau^1}=\c)=\sum_{k\geq0}q_k(x)$. Decomposing
over the two possible values of $l_1$ yields:
\[
q_k(x)=M^1(x,a)q_{k-1}(a)+M^1(x,b)q_{k-1}(b)\,.
\]
Therefore  the vector 
$\begin{pmatrix}
q_k(a)&q_k(b)
\end{pmatrix}$ satisfies the following recurrence relation:
\[
\begin{pmatrix}
  q_k(a)\\[\smalldim]q_k(b)
\end{pmatrix}
=N
\begin{pmatrix}
  q_{k-1}(a)\\[\smalldim]q_{k-1}(b)
\end{pmatrix}\,,\qquad\text{with }
N=
\begin{pmatrix}
  M^1(a,a)&M^1(a,b)\\[\smalldim]
M^1(b,a)&M^1(b,b)
\end{pmatrix}
\,.\]
We observe that $\begin{pmatrix}
  q_0(a)\\q_0(b)
\end{pmatrix}=\begin{pmatrix}
  M^1(a,\c)\\M^1(b,\c)
\end{pmatrix}$ and therefore:
\[
\begin{pmatrix}
  M^1_{a}(X^1_{\tau^1}=\c)\\[\smalldim]
  M^1_{b}(X^1_{\tau^1}=\c)
\end{pmatrix}=
(I-N)^{-1}\begin{pmatrix}
  M^1(a,\c)\\[\smalldim]
M^1(b,\c)
\end{pmatrix}\,.
\]
We find in a similar fashion:
\[
\begin{pmatrix}
  M^1_{a}(X^1_{\tau^1}=\d)\\[\smalldim]
  M^1_{b}(X^1_{\tau^1}=\d)
\end{pmatrix}=
(I-N)^{-1}\begin{pmatrix}
  M^1(a,\d)\\[\smalldim]
M^1(b,\d)
\end{pmatrix}\,,
\]
with same matrix~$N$. Finally we have:
\begin{equation}
\label{eq:31}
\begin{split}  M^1_{\c}(X^1_{\tau^1}=\c)&=M^1(\c,\c)+M^1(\c,a)M^1_a(X^1_{\tau^1}=\c)+M^1(\c,b)M^1_b(X^1_{\tau^1}=\c)\\
M^1_{\c}(X^1_{\tau^1}=\d)&=M^1(\c,\d)+M^1(\c,a)M^1_a(X^1_{\tau^1}=\d)+M^1(\c,b)M^1_b(X^1_{\tau^1}=\d)\,.
\end{split}
\end{equation}
And in a similar fashion:
\begin{equation}
\label{eq:44}
\begin{split}
  M^1_{\d}(X^1_{\tau^1}=\c)&=M^1(\d,\c)+M^1(\d,a)M^1_a(X^1_{\tau^1}=\c)+M^1(\d,b)M^1_b(X^1_{\tau^1}=\c)\\
M^1_{\d}(X^1_{\tau^1}=\d)&=M^1(\d,\d)+M^1(\d,a)M^1_a(X^1_{\tau^1}=\d)+M^1(\d,b)M^1_b(X^1_{\tau^1}=\d)\,.
\end{split}
\end{equation}
Applying these calculations to our numerical example, we find:
\begin{align*}
N&=
\begin{pmatrix}
\frac13&\frac13\\[\smalldim]
\frac12&\frac18  
\end{pmatrix}&
(I-N)^{-1}&=\frac{12}5
\begin{pmatrix}
  \frac78&\frac13\\[\smalldim]
\frac12&\frac23
\end{pmatrix}\\
\begin{pmatrix}
  M^1_a(X^1_{\tau^1}=\c)\\[\smalldim]
M^1_b(X^1_{\tau^1}=\c)
\end{pmatrix}&=
\frac{12}5\begin{pmatrix}
  \frac 13\\[\smalldim]
\frac14
\end{pmatrix}
&
\begin{pmatrix}
  M^1_a(X^1_{\tau^1}=\d)\\[\smalldim]
M^1_b(X^1_{\tau^1}=\d)
\end{pmatrix}&=\frac{1}5
\begin{pmatrix}
1\\[\smalldim]
2
\end{pmatrix}\,.
\end{align*}
We obtain thus, using Eqs.~(\ref{eq:31})(\ref{eq:44}):
\begin{align*}
  M^1_{\c}(X^1_{\tau^1}=\c)&=\frac{13}{20}&M^1_{\c}(X^1_{\tau^1}=\d)&=\frac7{20}\\
M^1_{\d}(X^1_{\tau^1}=\c)&=\frac{11}{20}&M^1_{\d}(X^1_{\tau^1}=\d)&=\frac9{20}
\end{align*}
Since we have taken $M^2=M^1$, we obtain the same laws depending on
the initial state $\c$ or $\d$ for $X^2_{\tau^2}$\,. The $2\times2$
transition matrix of $Y$ is now obtained by conditioning the free
product $(X^1_{\tau^1},X^2_{\tau^2})$ on
$X^1_{\tau^1}=X^2_{\tau^2}$\,, which yields the following transition
matrix:
\[
\begin{matrix}
  \c \\[\smalldim] \d
\end{matrix}
\begin{pmatrix}
  \frac{169}{218}&\frac{49}{218}\\[\smalldim]
\frac{121}{202}&\frac{81}{202}
\end{pmatrix}\,.
\]

\section*{Conclusion}
\label{sec:conclusion}

\subsection*{Summary of results}
\label{sec:summary-results}

Following the idea that, in a network, the knowledge a node has about
time is related to its local clock, and to its local clock only, we
have introduced a probabilistic model based on a simple trace model,
that allows private changes of states and synchronizations between two
sites. We have focused on a Markov model where local components are
independent up to the synchronization constraints, which brought us to
the formulation of a Markov property without reference to any time
index on the one hand, and to the Local Independence Property on the
other hand. Triples $(\Omega,\FFF,\bpr)$ where $(\Omega,\FFF)$ is the
space of trajectories and $\bpr$ is a probability measure satisfying
both properties have been constructed and entirely characterized by a
finite family of transition matrices, extending the familiar
transition matrix from discrete time Markov chain theory.

A singular feature of the model is the absence of \emph{constant
  times}; instead, only random times may be considered, and among them
stopping times play a distinguished role. Note that despite the
absence of a totally ordered time index, we can conduct probabilistic
reasoning about our two-components models at the level of stopping
times.


\subsection*{Potential applications}

Open research fields involving asynchronous systems are numerous. In
some cases, trace models have proved to be more relevant than
interleaving models: distributed observation, supervision and
diagnosis of concurrent systems, distributed optimization and planning
\cite{benveniste03:_diagn} provide examples.
In the formal verification community, people have considered
interleaving models for composing probabilistic systems (cf.\ the
discussion in the Introduction). Although product of Probabilistic
Automata for instance has shown to be efficient for developing proving
techniques based on bisimulation relations, it is worth trying other
ways for modeling network system where asynchrony plays an important
role.

One can therefore expect new advances in the theory of networked
systems through the development of a probabilistic layer for trace
models. In this respect, asymptotic analysis of probabilistic trace
models may have applications in network dimensioning.

\subsection*{Limitations and extensions}

Although the model of Markov concurrent process adopted in this paper
is limited to \emph{two} components only, it is important to notice
that it has a straightforward generalization to an arbitrary number
$n\geq2$ of components. In this generalized framework, the notion of
stopping time, the Asynchronous Strong Markov Property and all the
results developed in \S~\ref{sec:stopp-times-strong} carry over
without additional difficulty. The LIP may also be expressed for
$n\geq2$ components in a similar way than we did for two components
only.  However, the mere existence of Markov processes with $n\geq2$
components is not trivial to prove. This relies on the additional
combinatorial complexity that appears when at least four components
are involved, since then different synchronization events can occur
concurrently. Therefore the simple structure of trajectories given by
Proposition~\ref{prop:1} is no longer valid, making in turn the
constructions of this paper found in
Sections~\ref{sec:synchr-two-mark} and~\ref{sec:gener-constr-mark}
ineffective.

Nevertheless, the task of proving the existence of Markov processes
with the LIP has been tackled in~\cite{abbes11}, generalizing the
synchronization product of Markov chains. However, this construction
is not very natural, and its main advantage is to encourage further
study in this direction, since at least it ensures that the object of
study is not empty. 

Regarding a general theory of Markov multi-components processes, one
may retain the following elements from the present paper: firstly,
stopping times and the Asynchronous Strong Markov Property have a
straightforward extension to $n\geq2$ components. These are basic
tools that remain unchanged. Secondly, the generalized LIP allows to
focus on the synchronization process only, since it implies a
conditional decorrelation between the synchronization process on the
one hand, and the private parts of each component on the other hand.
The core of the remaining challenge is thus the construction and
characterization of the synchronization process---we have shown above
that, for two components, the synchronization process identifies with
a homogeneous Markov chain, a drastic simplification compared to the
general case of an arbitrary number of components. Recent work by
G.~Winskel~\cite{winskel13} on probabilistic event structures has
shown to be promising in this respect.

{\small
\section*{Acknowledgments}
\label{sec:aknowledgment}

Many thanks go to Albert Benveniste from IRISA in Rennes (France) for
his support, his help and his friendship.  I would like also to thank
the anonymous referees for their many comments and suggestions, and
the Editor Prakash Panangaden, to whom I am profoundly grateful.}

\bibliography{M2CP.bib}
\bibliographystyle{plain}

\vspace{-30 pt}
\end{document}